\pdfoutput=1 
\documentclass[12pt]{iopart-AG}
\usepackage{setstack}
\usepackage{graphicx,amssymb,amsthm,epsfig,amsbsy,amsfonts,mathrsfs}
\usepackage[english,francais]{babel}

\usepackage{wasysym}

\usepackage{url,hyperref}  

\usepackage{ulem}

\usepackage{xcolor}

\definecolor{Cblue}{HTML}{045FB4}
\definecolor{Cred}{HTML}{DF0101}
\definecolor{CDarkred}{HTML}{8A0808}

\hypersetup{
    colorlinks=true, 
    linktoc=all,     
    linkcolor=Cred,
    citecolor=Cblue
  }

  \definecolor{blue}{rgb}{0,0,1}
  \definecolor{green}{rgb}{0,.6,0}
  \definecolor{red}{rgb}{1,0,0}
  \definecolor{vio}{rgb}{1,0,1}
  \definecolor{uv}{rgb}{0.5,0,0.5}
  \definecolor{ama}{rgb}{0.3,0.3,0.3}

\definecolor{M_Beige}         {rgb}{0.96 , 0.96 , 0.86}

\definecolor{M_Brown}         {rgb}{0.65 , 0.16 , 0.16}

\definecolor{M_Gold}          {rgb}{1.00 , 0.84 , 0.00}

\definecolor{M_LemonChiffon}  {rgb}{1.00 , 0.98 , 0.80}

\definecolor{M_Orange}        {rgb}{1.00 , 0.60 , 0.00}

\definecolor{M_Pink}          {rgb}{0.80 , 0.55 , 0.60}

\definecolor{M_Violet}          {rgb}{0.83 , 0.21 , 0.93}

\definecolor{M_Green}          {rgb}{0.2 , 0.6 , 0.2}

\definecolor{M_Gray}          {rgb}{0.7 , 0.7 , 0.7}

\definecolor{M_BluPal}          {rgb}{0.7 , 0.7 , 0.9}

\renewcommand{\geq}{\geqslant}

\def\eqdef{\stackrel{\mbox{\tiny def}}{=}}     
\def\eqlaw{\stackrel{\mbox{\tiny (law)}}{=}}     
\newcommand{\ket}[1]{|\kern.3ex#1\kern.3ex\rangle}
\newcommand{\bra}[1]{\langle\kern.3ex #1 \kern.3ex|}
\newcommand{\scalar}[2]{\langle\kern.3ex{#1}\kern.3ex|\kern.3ex{#2}\kern.3ex\rangle}
\newcommand{\mean}[1]{\left\langle #1\right\rangle}
\newcommand{\smean}[1]{\langle #1\rangle}

\newcommand{\EXP}[1]{\mathrm{e}^{#1}}         

\newcommand{\im}{\mathop{\mathrm{Im}}\nolimits}      
\renewcommand{\tr}[1]{\mathop{\mathrm{tr}}\nolimits\left\{ #1 \right\}}  
\renewcommand{\min}[2]{\mathop{\mathrm{min}}\nolimits\left( #1 , #2\right)}

\def\I{{\rm i}}                  
\def\D{{\rm d}}                  

\newcommand{\deriv}[2]{\frac{\mathrm{d}#1}{\mathrm{d}#2}}
\newcommand{\derivp}[2]{\frac{\partial #1}{\partial #2}}

\newcommand\antiddots{\mathinner{\mkern2mu\raise1pt\hbox{.}\mkern2mu
    \newline \raise4pt\hbox{.}\mkern2mu\raise7pt\hbox{.}\mkern1mu}}

\def\Wt{\tau_\mathrm{W}}
\def\Ht{\tau_\mathrm{H}}
\def\Nc{N}
\def\Sm{\mathcal{S}}
\def\WSm{\mathcal{Q}}    
\def\invQ{\Gamma}      

\def\WSmA{\widetilde{\mathcal{Q}}}  

\def\Riccati{\mathcal{Z}}

\def\Im{\mathrm{Im}}

\def\O{\mathcal{O}}
\def\T{\mathrm{T}}
\def\Diag{\mathrm{Diag}}

\usepackage{wasysym}
\def\ItoDr{{\mathfrak{Drift}}}

\def\Sm{\mathcal{S}}
\def\Smt{\widetilde{\mathcal{S}}}
\def\WSmt{\widetilde{\mathcal{Q}}}    
\def\UR{\mathcal{U}_R}
\def\UL{\mathcal{U}_L}

\def\un{\mathbf{1}_\Nc}

\newcommand{\lambdabar}{{\mkern0.75mu\mathchar '26\mkern -9.75mu\lambda}}

\begin{document}

\renewcommand{\labelitemi}{$\bullet$}
\renewcommand{\labelitemii}{$\star$}

\selectlanguage{english}

\title[Wigner-Smith matrix and exponential functional of the matrix BM]{Wigner-Smith matrix, exponential functional of the matrix Brownian motion and matrix Dufresne identity}

\author{Aur\'elien Grabsch}
\address{Instituut-Lorentz,  Universiteit  Leiden,  P.O.  Box  9506,  2300  RA, Leiden, The Netherlands}

\author{Christophe Texier}
\address{LPTMS, CNRS, Universit\'e Paris-Saclay, 91405 Orsay cedex, France}

\begin{abstract}
We consider a multichannel wire with a disordered region of length $L$ and a reflecting boundary. 
The reflection of a wave of frequency $\omega$ is described by the scattering matrix $\mathcal{S}(\omega)$, encoding the probability amplitudes to be scattered from one channel to another.
The Wigner-Smith time delay matrix $\mathcal{Q}=-\mathrm{i}\, \mathcal{S}^\dagger\partial_\omega\mathcal{S}$ is another important matrix, which encodes temporal aspects of the scattering process.
In order to study its statistical properties, we split the scattering matrix in terms of  two unitary matrices, 
$\mathcal{S}=\mathrm{e}^{2\mathrm{i}kL}\mathcal{U}_L\mathcal{U}_R$ (with $\mathcal{U}_L=\mathcal{U}_R^\mathrm{T}$ in the presence of time reversal symmetry), and introduce a novel symmetrisation procedure for the Wigner-Smith matrix~:
$\widetilde{\mathcal{Q}}
=\mathcal{U}_R\,\mathcal{Q}\,\mathcal{U}_R^\dagger
= (2L/v)\,\mathbf{1}_N
-\mathrm{i}\,\mathcal{U}_L^\dagger\partial_\omega\big(\mathcal{U}_L\mathcal{U}_R\big)\,\mathcal{U}_R^\dagger$,
where $k$ is the wave vector and $v$ the group velocity.
We demonstrate that $\widetilde{\mathcal{Q}}$ can be expressed under the form of an exponential functional of a matrix Brownian motion. 
For semi-infinite wires, $L\to\infty$, using a matricial extension of the Dufresne identity, we recover straightforwardly the joint distribution for $\mathcal{Q}$'s eigenvalues of Brouwer and Beenakker [Physica E \textbf{9}, 463 (2001)].
For finite length $L$, the exponential functional representation is used to calculate the first moments
$\langle\mathrm{tr}(\mathcal{Q})\rangle$, 
$\langle\mathrm{tr}(\mathcal{Q}^2)\rangle$ and $\langle\big[\mathrm{tr}(\mathcal{Q})\big]^2\rangle$.
Finally we derive a partial differential equation for the resolvent 
$g(z;L)=\lim_{N\to\infty}(1/N)\,\mathrm{tr}\big\{\big( z\,\mathbf{1}_N - N\,\mathcal{Q}\big)^{-1}\big\}$ 
in the large $N$ limit. 

%
%
\end{abstract} 







\maketitle



\section{Introduction}
\label{sec:Intro}

Scattering of waves in complex media has been the subject of intense investigations for several decades, with applications in many areas of physics, ranging from compound-nucleus reactions \cite{VerWeiZir85,MitRicWei10}, chaotic billiards \cite{GuhMulWei98}, electromagnetic waves in random media \cite{SebGen98} to coherent electronic transport \cite{Bee97,MelKum04}.
When the wave is elastically scattered by the static potential, the scattering process is encoded in the on-shell scattering matrix, with elements $\Sm_{ab}(\omega)$ characterizing the amplitude of the wave in the scattering channel $a$, if a wave of frequency $\omega$ was injected in channel $b$ (channels can be the open transverse modes of some wave guides).
Given the $\Nc\times\Nc$ scattering matrix $\Sm$ as a function of the frequency, it is possible to construct another important matrix, known as the Wigner-Smith time delay matrix~\cite{Wig55,Smi60} 
\begin{equation}
  \label{eq:DefWSm}
  \WSm=-\I\,\Sm^\dagger\,\partial_\omega\Sm
\end{equation}
encoding several sets of times characterizing the scattering process (cf. Refs. \cite{CarNus02} and \cite{Tex16} and references therein for a recent review of these concepts).

In complex media, it is natural to investigate the statistical properties of these two matrices (their sample to sample fluctuations).
The most studied setting is a chaotic cavity, i.e. a \textit{zero-dimensional} situation. 
In such devices, the wave is injected through wave guides.
The complex nature of the dynamics inside the cavity leads to a random matrix formulation based on some maximum entropy principle \cite{MelPerSel85,MelKum04}. For example, assuming perfect contacts, it is natural to assume that $\Sm$ belongs to one of the circular ensembles (COE, CUE or CSE), depending on the presence or absence of time reversal symmetry and/or spin rotational symmetry \cite{Meh04}.
Based on such assumptions, with additional modelling of the frequency dependence \cite{BroBut97}, the distribution of the Wigner-Smith matrix eigenvalues has been obtained by Brouwer, Frahm and Beenakker (BFB)~\cite{BroFraBee97,BroFraBee99}. 
Precisely, introducing the symmetrised Wigner-Smith matrix 
\begin{equation}
\label{eq:DefSymWSm}
  \WSm_s  = -\I\, \Sm^{-1/2} \derivp{\Sm}{\omega}\, \Sm^{-1/2} 
  \:,
\end{equation}
its inverse $\invQ=\Ht\,\WSm_s^{-1}$, where $\Ht$ is the Heisenberg time,~\footnote{$\Ht=2\pi/\delta\omega$ where $\delta\omega$ is the mean level spacing between eigenmodes of the cavity.} was shown to be distributed according to a specific instance of the Laguerre ensemble of random matrix theory, 
$P(\Gamma)\propto\big(\det\Gamma\big)^{\beta\Nc/2}\,\EXP{-(\beta/2)\tr{\Gamma}}$, over the set of Hermitian matrices with positive eigenvalues.
$\beta$ is the Dyson index ($\beta=1$ when time reversal symmetry holds and $\beta=2$ if not).
Based on this distribution, many results have been obtained for ideal contacts~:
cumulants \cite{MezSim13} and distribution \cite{TexMaj13} of its trace $\tr{\WSm}$, or other correlations \cite{MezSim11,MezSim12,GraTex15,Cun15,CunMezSimViv16} (see the updated preprint version of Ref.~\cite{Tex16} for an exhaustive review).
Several generalizations of BFB's distribution have been obtained more recently~:
the case of non-ideal contacts has been studied~\cite{MarSchBee16,GraSavTex18}, BdG symmetry classes \cite{MarSchBee16} and the effect of absorption (for ideal contacts)~\cite{Gra20}. 

Several results are also known beyond the zero-dimensional case.
The case of a strictly one-dimensional disordered wire of length $L$ with a reflecting boundary, corresponding to $\Nc=1$ scattering channel, is best understood. 
In this case the wave is expected to be localised by the disorder, on a typical scale $\xi$ (the localisation length). 
A wave packet may remain trapped a long time if the localisation center is far from the edge, which gives rise to narrow resonances.
In the high energy/weak disorder regime, when universality is expected, the relation to universality of localisation properties was established, which has led to derive a representation of the Wigner time delay  under the form of the exponential functional of the Brownian motion (BM)~\cite{TexCom99} \footnote{An identity in law relates two quantities with same statistical properties. For example, the well-known scaling property of the Brownian motion can be written $B(\lambda x)\eqlaw \sqrt{\lambda}\,B(x)$.}
\begin{equation}
  \label{eq:WTDfBM}
   \WSm \eqlaw 2\tau_\xi \int_0^{L/\xi} \D x \,  \EXP{-2x + 2B(x)}
\end{equation} 
where $B(x)$ is a normalised Brownian motion, such that $\mean{B(x)}=0$ and $\mean{B(x)B(x')}=\min{x}{x'}$
(in Refs.~\cite{FarTsa94,ComTex97} a different form, although equivalent to \eref{eq:WTDfBM}, was proposed).
The characteristic scale
\begin{equation}
 \tau_\xi=\xi/v
\end{equation} 
is the time needed by the particle with group velocity $v$ to cover the localisation length~$\xi$.
Using known results on exponential functionals of the BM \cite{MonCom94,ComMonYor98}, the representation \eref{eq:WTDfBM} has allowed to derive the moments \cite{TexCom99,ComTex97} of $\WSm$ and its full distribution $P_L(\tau)$ \cite{ComTex97} for finite $L$.
The limit law of the Wigner time delay, for $L \to \infty$, was derived in Refs.~\cite{ComTex97,TexCom99} and also in \cite{OssKotGei00} within a tight binding model~\footnote{See the arXiv version of Ref.~\cite{Tex16} for a detailed review.}
\begin{equation}
  \label{eq:LimitLawComtetTexier}
  P_\infty(\tau) = \frac{\tau_\xi}{\tau^2}\,\EXP{-\tau_\xi/\tau}
  \:.
\end{equation}

The exponential functional of the Brownian motion
\begin{equation}
  \label{eq:EFBM}
  Z^{(\mu)}_L = \int_0^{L} \D x \, \lambda(x)^2
  \hspace{0.5cm}\mbox{with }   \lambda(x) = \EXP{-\mu\, x + B(x)}
\end{equation}
with other functionals have attracted a considerable interest in the mathematical community \cite{Yor00,MatYor05a,MatYor05b}~; the relation with several physical problems is reviewed in Refs.~\cite{Mon95,ComDesTex05}. 
They have also found several applications in mathematical finance, in the context of which Dufresne has obtained the remarkable identity \cite{Duf90}
\begin{equation}
  \label{eq:Dufresne}
  Z^{(\mu)}_\infty \eqlaw \frac{1}{\gamma^{(\mu)}}  \hspace{0.5cm}\mbox{for } \mu>0
  \:,
\end{equation}
where  $\gamma^{(\mu)}$ obeys the Gamma-law
\begin{equation}
  p(\gamma) =\frac{1}{2^\mu\Gamma(\mu)}\, \gamma^{\mu-1} \,\EXP{-\gamma/2}
  \:.
\end{equation}
The representation \eref{eq:EFBM} makes clear that \eref{eq:WTDfBM} corresponds to a drift $\mu=1$. Hence the limit law  \eref{eq:LimitLawComtetTexier} is a direct consequence of the Dufresne identity \eref{eq:Dufresne}.

Beyond the weak disorder/high energy universal regime in one dimension, some other results have been obtained.
In the strictly one dimensional case, various results were also derived in the low energy/strong disorder regime \cite{TexCom99,Tex99}.~\footnote{
  A non-trivial distribution for the time delay for the dimer model with delocalisation points \cite{DunWuPhi90} 
  was also obtained in chapter 6 of~\cite{Tex99}.
} 
The case of higher dimensions has also been investigated~\cite{OssFyo05} (see the review article \cite{Kot05}).
More recently, the marginal distribution of the proper time delays was studied by Ossipov~\cite{Oss18}, claiming to describe also the metallic regime in $d>2$~; we criticize this statement at the end of the paper.

Another interesting case, which is more tractable, is the intermediate situation of quasi-one-dimensional systems, i.e. multichannel disordered wires. 
The assumption that channels are statistically equivalent (isotropy) allows to derive analytical results, such as the Lyapunov spectrum or the statistics of transmission probabilities \cite{Bee97}.
The joint distribution of eigenvalues of $\invQ=\tau_\xi\,\WSm^{-1}$ 
has been derived for a semi-infinite disordered wires in Refs.~\cite{Bee01,BeeBro01}
\begin{equation}
  \label{eq:BeenakkerBrouwer2001}
   \mathcal{P}_\Nc(\gamma_1,\ldots,\gamma_N) \propto \prod_{i<j} |\gamma_i-\gamma_j|^\beta
   \prod_n\EXP{-\gamma_n/2}
   \hspace{0.5cm}\mbox{for }L\to\infty
   \:,
\end{equation}
corresponding to the matrix 
 distribution 
\begin{equation}
  \label{eq:LagWishDist}
  P(\Gamma) \propto \EXP{-(1/2)\tr{\Gamma}}
  \:,
\end{equation}
defined over the set of Hermitian matrices with positive eigenvalues.
For $\Nc=1$, the distribution corresponds to \eref{eq:LimitLawComtetTexier}.
 This is a different instance of the Laguerre ensemble of random matrix theory than the one obtained for chaotic quantum dots. 
This result has been used to show that the distribution of the Wigner time delay, i.e. the trace $\Wt=(1/\Nc)\tr{\WSm}$, becomes independent of $\Nc$ in the large $\Nc$ limit \cite{GraTex16b}~:
\begin{equation}
  \label{eq:GrabschTexier2016}
  \hspace{-1cm}
   \mathscr{P}^{(\beta)}_\Nc(\tau)  
   \simeq 
   \frac{A_\beta}{\tau^2}
   \exp\left\{
     -\frac{27\tau_\xi^2}{64\beta\,\tau^2} +\left(\frac{2}{\beta}-1\right)\frac{9(2-\sqrt{3})\tau_\xi}{8\,\tau}
   \right\}
   \hspace{0.5cm}\mbox{for }L\to\infty
\end{equation}
where $A_\beta$ is a normalisation.
This shows in particular that, as in the $\Nc=1$ channel case, all moments $\smean{\tr{\WSm}^n}$ are \textit{infinite} for $L\to\infty$.
The physical origin of the divergence lies in the proliferation of very narrow resonances (this is discussed for the case $\Nc=1$ in \cite{TexCom99}).
Much less is known for finite length $L$. 
Using the fact that $\Nc\Wt/(2\pi)=(2\pi)^{-1}\tr{\WSm}$ can be interpreted as the density of states of the open system (see~\ref{app:FSrelation}), we can write 
\begin{equation}
  \label{eq:MeanTraceQ}
   \smean{\tr{\WSm}}=\frac{\Nc L}{k}
   \:,
\end{equation}
nevertheless the behaviour of higher moments is an open question.
It is the aim of the present article to study this problem and provide some statistical information on the Wigner-Smith time delay matrix for disordered wire of finite length~$L$.
For this purpose, we will obtain a generalisation of the representation \eref{eq:WTDfBM}, for $\Nc>1$.
In particular, this will provide a straightforward derivation of the distribution \eref{eq:BeenakkerBrouwer2001} for $L\to\infty$, by using an extension of the Dufresne identity \eref{eq:Dufresne} to the multichannel case, when $\WSm$ is a $\Nc\times\Nc$ random matrix. 
Furthermore, this will allow a determination of the moments.


\subsection{The model of multichannel  disordered wires}
\label{Subsec:model}

The (quasi-one-dimensional) multichannel model under investigation in the article is the Schr\"odinger equation 
\begin{equation}
  \label{eq:SchrodEq}
  H\psi(x)=\varepsilon\,\psi(x)
  \hspace{0.5cm}\mbox{with}\hspace{0.5cm}
  H = -\mathbf{1}_\Nc \, \partial_x^2 + V(x)
  \:,
\end{equation}
 where $\psi(x)$ is a column vector with $\Nc$ components, coupled by the potential $V(x)$.
We consider the case where $V(x)$ is a  $\Nc\times\Nc$ matrix Gaussian white noise with zero mean and correlations
\begin{equation}
  \label{eq:Ciso}
  \mean{ V_{ab}(x) V_{cd}^*(x')} = \sigma\, C_{ab,cd}\, \delta(x-x')
  \:,
\end{equation}
where $\sigma$ is the disorder strength (with dimension $[\sigma]=L^{-3}$). 
We will assume \textit{isotropy} among the channels,~\footnote{
  (\ref{eq:Ciso},\ref{eq:IsotropyAssumption}) correspond to the distribution
  $P[V(x)]=\exp-\frac{1}{2\sigma}\int\D x\,\tr{V(x)^2}$.
} i.e. the invariance of the statistical properties of $V(x)$ under orthogonal ($\beta=1$) or unitary ($\beta=2$) transformations.
This leads to the correlations between channels
\begin{equation}
  \label{eq:IsotropyAssumption}
  \hspace{-2cm}
  C_{ab,cd}
  = \frac{\beta}{2}\,\delta_{ac} \delta_{bd}+\left(1-\frac{\beta}{2}\right)\,\delta_{ad} \delta_{bc}
  = 
  \left\lbrace
    \begin{array}{ll}
      \frac{1}{2}
      \left( \delta_{ac} \delta_{bd} + \delta_{ad} \delta_{bc} \right)
      & \text{for } \beta = 1 \mbox{ (TRS)} \:,
      \\[0.2cm]
      \delta_{ac} \delta_{bd}
      & \text{for } \beta = 2 \mbox{ (no TRS).}
  \end{array}
  \right.
\end{equation}
An important scale of the problem is the elastic scattering rate $1/\tau_e$, related to the self energy $\Sigma^\mathrm{R}_{ab}$ by~:
\begin{equation}
  \frac{1}{2\tau_e} = - \Im \Sigma^\mathrm{R}_{aa}
  \:.
\end{equation}
We denote by 
$G^\mathrm{R}_{a,b}(x,x')=\delta_{a,b}\,\frac{1}{2\I k}\EXP{\I k|x-x'|}$ the free retarded Green's function for energy $\varepsilon=k^2$.
In the weak disorder regime we have 
\begin{equation}
  \Sigma^\mathrm{R}_{ab} \simeq \sum_c \sigma\,C_{ac,bc} \,G^\mathrm{R}_{c,c}(0,0)
  = - \frac{\I\sigma}{2k}\sum_c C_{ac,bc}
\end{equation}
We deduce the elastic mean free path $\ell_e=v\tau_e$, where $v=\partial \varepsilon/\partial k=2k$ is the group velocity, in terms of the disorder strength $\sigma$ 
\begin{equation}
  \ell_e = \frac{2k^2}{\mu\,\sigma}
\end{equation}
where
\begin{equation}
  \label{eq:DriftLoc}
   \boxed{
  \mu  = \sum_{b}  C_{ab,ab} = 1 + \frac{\beta}{2}(\Nc-1)
  }
\end{equation}
As we will see, for weak disorder $\varepsilon=k^2\gg\sigma^{2/3}$,  the localisation length is given by  
\begin{equation}
  \boxed{
  \xi = \frac{8k^2}{\sigma}
  = 4\mu\, \ell_e = 2 \left[2 + \beta\left(\Nc-1\right)\right] \ell_e
  }
\end{equation}
which is the well-known dependence in $\beta$ and $\Nc$ (obtained within a different model in Ref.~\cite{Bee97}).
  
Below, we study the scattering problem on $\mathbb{R}^+$ for a potential defined on $[0,L]$ (and vanishing outside the interval).
The eigenstate corresponding to inject the wave in channel $b\in\{1,\ldots,\Nc\}$ is a $\Nc$-component vector denoted $\psi_\varepsilon^{(b)}(x)$. 
We write the $a$-th component, i.e. the amplitude in channel $a$, in the free region as
\begin{equation}
  \label{eq:SSS}
  \left[\psi_\varepsilon^{(b)}(x)\right]_a  =
  \frac{1}{\sqrt{hv}}
  \left(
    \delta_{ab}\, \EXP{-\I k (x-L)} + \Sm_{ab}(\varepsilon)\,\EXP{\I k (x-L)} 
  \right)
  \hspace{0.5cm}\mbox{for } x\geq L
  \:.
\end{equation}
The prefactor ensures the normalisation \cite{Tex15book} 
$
\scalar{\psi_{\varepsilon}^{(a)}}{\psi_{\varepsilon'}^{(b)}} =\delta_{ab}\,\delta(\varepsilon-\varepsilon')
$.


\subsection{Statement of the main results}

Our analysis is based on a new symmetrization procedure of the Wigner-Smith matrix.
Assuming that all channels are controlled by the same wave vector $k$ in the absence of disorder, we extract rapid oscillations of the scattering matrix $\Sm=\Smt\,\EXP{2\I k L}$, where $\Smt$ is controlled by slow variables. 
We use a \og square root trick \fg{} in order to decompose it in terms of two unitary matrices $\Smt=\UL\mathcal{U}_R$.
In the presence of TRS ($\beta=1$), $\UL=\UR^\T$ ensures the property $\Smt=\Smt^\T$.
In the absence of TRS ($\beta=2$), they are chosen such that they obey two matrix stochastic differential equations (SDE) of convenient form.
Then, the Wigner-Smith matrix is symmetrised as 
\begin{equation*}
  \label{eq:MRtheunitarytransf}
  \WSmA
  =\UR\,\mathcal{Q}\,\UR^\dagger
  =\frac{2L}{v}\,\un-\I\,\UL^\dagger\,\partial_\varepsilon\big(\UL\UR\big)\,\UR^\dagger
  \:,
\end{equation*}
where $v$ is the group velocity. The first term is the result in the absence of the disorder~: $2L/v$ is the time needed to go back and forth in the sample when $V=0$.
Our analysis relies on the decoupling between fast and slow variables in the high energy/weak disorder regime and on an isotropy assumption (invariance under exchange of channels). 
One of our main result is the matrix SDE 
\begin{equation*}
  \label{eq:MSDE-WSm-mr}
  \derivp{}{L} \WSmt = \frac{\un}{k}
  - \frac{2 \mu}{\xi} \WSmt
  + \frac{1}{\sqrt{\xi}} \left( \WSmt\, \eta(L) + \eta(L) \, \WSmt \right)
  \hspace{1cm} \text{(Stratonovich)}  
  \:,
\end{equation*}
for $  \mu = 1 + \frac{\beta}{2}(\Nc-1)$ and where $\eta(x)$ is a normalised Hermitian Gaussian white noise, 
$\mean{\eta_{ab}(x)\eta_{cd}^*(x')}=C_{ab,cd}\,\delta(x-x')$ with \eref{eq:IsotropyAssumption}.
From this matrix SDE,
we deduce a representation of the Wigner-Smith time delay matrix under the form of an exponential functional of a \textit{matrix} Brownian motion. 
\begin{equation}
  \label{eq:MainResult}
  \WSmA \eqlaw 2 \tau_\xi
  \int_0^{L/\xi}\D x\, \Lambda(x)^\dagger\,\Lambda(x)
\end{equation}
where $\Lambda(x)$ obeys the matrix SDE
\begin{equation*}
  \partial_x \Lambda(x) = -\mu\, \Lambda(x)  + \eta(x)\, \Lambda(x)
  \hspace{1cm} \text{(Stratonovich)}  
  \:,
\end{equation*}
with $\Lambda(0)=\un$.
We may also write
\begin{equation}
  \label{eq:LambdaChronological}
  \Lambda(x) = \mathrm{T}\,\EXP{-\mu x + \int_0^x\D t\,\eta(t)}
\end{equation}
where $\mathrm{T}$ denotes chronological ordering, to make the contact with formulae (\ref{eq:WTDfBM},\ref{eq:EFBM}) more explicit.
For $\Nc=1$ channel, we recover \eref{eq:WTDfBM} (i.e. the form \eref{eq:EFBM} for $\mu=1$).
The representation \eref{eq:MainResult} has allowed us to recover straightforwardly the result of Beenakker and Brouwer \eref{eq:BeenakkerBrouwer2001}, by using a matrix generalization of the Dufresne identity
\begin{equation*}
  \label{eq:MatrixDufresneAlaGrabsch}
    \WSmA \eqlaw 2\tau_\xi \,\Gamma^{-1}
     \hspace{0.5cm}\mbox{for } L\to\infty
\end{equation*}
where $\Gamma$ obeys the Laguerre distribution \eref{eq:LagWishDist}.  

As an application of the matrix SDE for $\WSmt$, we show on the example of $\mean{\tr{\WSm}}$, $\mean{\tr{\WSm}^2}$ and $\mean{\tr{\WSm^2}}$, how moments can be computed.

Finally, we reconsider the problem studied by Ossipov \cite{Oss18}, within our model based on \textit{isotropy} assumption. 
We recover Ossipov's equation for the resolvent 
$g(z;L)=\lim_{N\to\infty}(1/\Nc)\,\mathrm{tr}\big\{\big[ z\,\mathbf{1}_N - N\,\WSm/(2\tau_\xi)\big]^{-1}\big\}$, which casts doubts on Ossipov's claim to describe the metallic phase in dimension $d>2$, as our model describes disordered wires transversally ergodic.


\subsection{Outline}

In Section~\ref{sec:WSlocalisation}, starting from a representation of the Wigner-Smith matrix in terms of the wave function, we show that localisation properties in multichannel disordered wires explain the origin of the relation with exponential functional of the matrix Brownian motion.
The following sections are devoted to a more precise derivation of this relation, with no prior knowledge of the localisation properties.
The analysis is based on the study of matrix stochastic differential equations (MSDE)~: the main SDE are derived Section~\ref{sec:SDEs}.
Then, Section~\ref{sec:WeakDisEff} discusses the elimination of fast variables in the high energy regime, leading to new MSDE for slow variables.
A new symmetrisation procedure of the Wigner-Smith matrix is introduced in Section~\ref{sec:NewSym}.
The isotropic assumption is introduced in Section~\ref{sec:IsoCase}, which allows, together with the new symmetrisation, the decoupling of the scattering matrix and the symmetrised Wigner-Smith matrix, leading eventually to the representation as an exponential functional of the matrix Brownian motion.
The relation with the matricial generalization of the Dufresne identity is discussed in Section~\ref{sec:MatrixDufresne}.
The representation is used in Section~\ref{sec:Moments} in order to derive the first moments for finite length.
Finally,  in Section~\ref{sec:MarginalDensity}, we discuss the resolvent of the Wigner-Smith matrix in the large $\Nc$ limit, i.e. the Stieltjes transform of the density of eigenvalues.


\section{Wigner-Smith matrix, localisation and exponential functional of the BM}
\label{sec:WSlocalisation}

This section presents some (partly heuristic) arguments explaining the origin of our main result, Eq.~\eref{eq:MainResult}, from the localisation properties in multichannel disordered wires. 
The model under investigation in the article, introduced in Subsection~\ref{Subsec:model}, 
is the Schr\"odinger equation \eref{eq:SchrodEq} for a $\Nc$ component wave function. 
We study here the scattering problem, i.e. eigenstates of the form \eref{eq:SSS}. 
For a given energy $\varepsilon$, we can construct $\Nc$ independent solutions  $\{\psi^{(a)}(x)\}_{a=1,\ldots,\Nc}$, corresponding to inject the incoming wave in one of the $\Nc$ channels.
The study of these $\Nc$ solutions can be \og parallelised \fg{} if we gather the $\Nc$ independent column vectors in the matrix wave function
\begin{equation}
  \Psi_\varepsilon (x) 
  = 
  \left(
  \begin{array}{ccc}
    \psi^{(1)}(x) & \cdots & \psi^{(\Nc)}(x)
  \end{array}
  \right)
\end{equation}
which behaves, in the disorder free region, as  
\begin{equation}
  \label{eq:AsymptoticPhi}
  \Psi_\varepsilon(x) = 
  \frac{1}{\sqrt{4\pi k}}
  \left(
    \mathbf{1}_\Nc\,\EXP{-\I k (x-L)} + \Sm(\varepsilon)\, \EXP{\I k (x-L)} 
  \right)
  \hspace{0.5cm}\mbox{for } x\geq L
  \:.
\end{equation}
The solution obeys the Schr\"odinger equation
\begin{equation}
  \label{eq:MatrixSchrod}
  -\Psi_\varepsilon''(x) + V(x)\, \Psi_\varepsilon(x) = \varepsilon\, \Psi_\varepsilon(x)
  \hspace{0.5cm}\mbox{for }
  x\geq0
  \:.
\end{equation}
As shown in \ref{app:FSrelation}, the wave function matrix is related to the Wigner-Smith matrix by the exact relation
\begin{equation}
  \label{eq:MatrixKreinFriedel}
    \boxed{
    \int_0^L\D x\, \Psi_\varepsilon^\dagger(x) \Psi_\varepsilon(x)
    = 
    \frac{1}{2\pi}\left(
      \WSm
      + \frac{\Sm-\Sm^\dagger}{4\I\varepsilon }
    \right)
    }
\end{equation}
which assumes Dirichlet boundary conditions $\Psi_\varepsilon(0) = 0$.

Eq.~\eref{eq:MatrixKreinFriedel} allows to understand easily the origin of the relation between the Wigner-Smith matrix and exponential functionals of the BM~; we follow and extend the argument given in Ref.~\cite{TexCom99} for the case $\Nc=1$. 
In the high energy/weak disorder regime, we can neglect the last term of \eref{eq:MatrixKreinFriedel} and write
\begin{equation}
  \label{eq:StartingPointSection}
  \WSm \simeq 2\pi\int_0^L\D x\, \Psi_\varepsilon^\dagger(x) \Psi_\varepsilon(x)
  \:.
\end{equation}
The wave function $\Psi_\varepsilon(x)$ presents fast oscillations on the scale $k^{-1}$ while its envelope is a smooth function, damped over scales given by the \textit{Lyapunov spectrum}.

For $\Nc=1$ (strictly one-dimensional case), 
we recall the argument of Ref.~\cite{TexCom99} leading to the representation~\eref{eq:WTDfBM}~:
the wave function in the disordered region may be parametrised as 
$\psi(x)=\frac{1}{\sqrt{\pi k}}\big[\varphi(x)/\varphi(L)\big]\,\sin\theta(x)$ where $\varphi(x)$ is an envelope and $\theta(x)$ a phase which controls the rapid oscillations. 
The presence of $\varphi(L)^{-1}$ ensures the matching on the behaviour \eref{eq:SSS}.
In the integral $\int_0^L\D x\,|\psi(x)|^2$, one can average over the fast oscillations, which corresponds to perform $\psi(x)\longrightarrow\frac{1}{\sqrt{2\pi k}}\varphi(x)/\varphi(L)$ in the integral.
The growth of the envelope is controlled by the Lyapunov exponent $\gamma$, inverse localisation length $\xi=1/\gamma$~: it is known to obey the SDE $\varphi'(x)=[\gamma+\sqrt{\gamma}\,\eta(x)]\varphi(x)$ \cite{AntPasSly81}, where $\eta(x)$ is a normalised Gaussian white noise (the fact that the diffusion and the drift are equal is known as \og single parameter scaling \fg{} \cite{CohRotSha88}~; see the recent broader discussion \cite{Tex19b}). 
A change of variable $x\to L-x$ in the integral eventually leads to the representation~\eref{eq:WTDfBM}.

We now extend the argument to the multichannel case. 
Let us now assume that averaging over the fast oscillations of the matrix wave function corresponds to perform a similar substitution
\begin{equation}
  \label{eq:SubstitutionPsiPhi}
  \Psi_\varepsilon(x) \longrightarrow \frac{1}{\sqrt{2\pi k}}\,\Phi(x)\,\Phi(L)^{-1}
\end{equation}
in \eref{eq:StartingPointSection}, where $\Phi(x)$ describes the smooth evolution of the envelope of the wave function. 
It is expected to obey the MSDE
\begin{equation}
  \label{eq:MSDEPhi}
  \partial_x\Phi(x) = \left( \tilde{\mu}\,D + \sqrt{D}\,\eta(x) \right) \Phi(x)
\end{equation}
where $\eta(x)$ a normalised $\Nc\times\Nc$ matrix Gaussian white noise. 
The drift $\tilde{\mu}$ and the diffusion constant $D$ can be related to the well-known localisation properties from the three following remarks~:
\begin{itemize}
\item
  The Lyapunov spectrum of $X'(x)=\eta(x)\,X(x)$ in the orthogonal case has been obtained by Le Jan \cite{Lej85} and Newman \cite{New86}~:
  $\gamma_n = \frac{\beta}{2}\left( \Nc-2n+1\right)$ for $n\in\{1,\ldots,\Nc\}$ (for the unitary case, cf.~\cite{Gra18}).
  Thus \eref{eq:MSDEPhi} is related to the Lyapunov spectrum 
  $
  \gamma_n = D\Big[\tilde{\mu}+(\beta/2) ( 2n-1 - \Nc )\Big]
  $
  for 
  $n\in\{1,\ldots,\Nc\}$.
\item
  The Lyapunov spectrum characterizing localisation in multichannel disordered wires is known \cite{Bee97} 
  $ \gamma_n \propto 1 + \beta\, (n-1) $.
\item
  The $\Nc=1$ case coincides with the striclty one dimensional Lyapunov exponent $\gamma_1 = \frac{\sigma}{8k^2}$ (for high energy) \cite{AntPasSly81}.
\end{itemize}
The three remarks lead to $D=\sigma/(8k^2)$ and $\tilde{\mu} = 1 + \frac{\beta}{2}(\Nc-1)\equiv\mu$, coincinding with the drift introduced above, Eq.~\eref{eq:DriftLoc}. Thus, the Lyapunov spectrum for $\Phi(x)$ (i.e. for the wave function $\Psi_\varepsilon(x)$) is
\begin{equation}
  \gamma_n =  \frac{\sigma}{8k^2}
  \left( 1 + \beta \, (n-1) \right) 
    \hspace{0.5cm}\mbox{for }
    n\in\{1,\ldots,\Nc\}
    \:.
\end{equation}
The localisation length is given by the smallest Lyapunov exponent
\begin{equation}
  \xi = \frac{1}{\gamma_1} = \frac{8k^2}{\sigma}
  \:.
\end{equation}

The substitution \eref{eq:SubstitutionPsiPhi} leads to 
\begin{equation}
  \label{eq:WSinterm}
   \WSm \overset{\mathrm{conjecture}}{=} \frac{1}{k}
   \int_0^{L}\D x\, \left(\Phi(L)^\dagger\right)^{-1}\Phi(x)^\dagger\,\Phi(x)\,\Phi(L)^{-1}
\end{equation}
(remind that \eref{eq:SubstitutionPsiPhi} has not been fully justified).
The change of variable  
$\Lambda(x/\xi)=\Phi(L-x)\,\Phi(L)^{-1}$, allows to rewrite the functional as
\begin{equation}
  \label{eq:PlausibleMainResult}
  \hspace{-1cm}
   \WSm \overset{\mathrm{conjecture}}{=}
   2\tau_\xi
   \int_0^{L/\xi}\D x\, \Lambda(x)^\dagger\,\Lambda(x)
  \hspace{0.5cm}\mbox{where}\hspace{0.5cm}
  \partial_x\Lambda(x) = \left( -\mu + \eta(x) \right) \Lambda(x)
\end{equation}
for $\Lambda(0)=\un$.
The scale is $2\tau_\xi=\xi/k$.
The matrix Dufresne identity states that 
\eref{eq:PlausibleMainResult} 
has a limit law for $L\to\infty$~:
precisely, $\Gamma\eqlaw\big(\int_0^\infty\D x\, \Lambda(x)^\dagger\,\Lambda(x)\big)^{-1}$ is distributed according to the Wishart distribution
\begin{equation}
  \label{eq:WishartOfRiderValko}
  P(\Gamma) = \mathcal{C}_{\Nc,\beta}\,
  (\det \Gamma)^{\mu-1-\beta(\Nc-1)/2}
  \EXP{-(1/2)\tr{\Gamma}}
  \hspace{0.5cm}\mbox{for } \mu>\frac{\beta}{2}(\Nc-1)
  \:,
\end{equation}
which is proven in Section~\ref{sec:MatrixDufresne} (and for $\beta=1$ in Ref.~\cite{RidVal15}).
The distribution is defined over the set of positive Hermitian matrices, $\Gamma>0$, i.e. matrices with positive eigenvalues. $\mathcal{C}_{\Nc,\beta}$ is a normalisation constant. 
Using \eref{eq:DriftLoc} we recover the distribution \eref{eq:LagWishDist}. 

The above derivation makes clear the relation between the statistical properties of the Wigner-Smith matrix and localisation properties, which emphasizes their universal character.
However the argumentation of this section has a weakness~:
the substitution \eref{eq:SubstitutionPsiPhi} is a rather strong assumption. 
Adding a unitary matrix $U(x)$, controlled by slow variables, to the wave function would not change the Lyapunov spectrum, however the substitution $\Psi_\varepsilon(x)\longrightarrow(2\pi k)^{-1/2}\,\Phi(x)\,U(x)\,U(L)^{-1}\,\Phi(L)^{-1}$, would not lead to~\eref{eq:PlausibleMainResult}. 
In the next sections, we follow a more rigorous approach based on the analysis of matrix SDE, which generalizes to the mulichannel case the method of Ref.~\cite{ComTex97} for $\Nc=1$.
We will show that the \textit{symmetrised} Wigner-Smith matrix admits the representation~\eref{eq:PlausibleMainResult}.


\section{Matrix stochastic differential equations for $\Sm$ and $\WSm$}
\label{sec:SDEs}

In this section, we derive the main matrix stochastic differential equation (MSDE)  for the scattering matrix and the Wigner-Smith matrix, at the heart of our analysis. 
A convenient starting point is to  introduce the Riccati matrix 
\begin{equation}
  Z(x) = \Psi'(x) \Psi(x)^{-1}
\end{equation}
(we drop the label $_\varepsilon$ in the wave function).
From \eref{eq:MatrixSchrod}, it is straightforward to get
\begin{equation}
  \label{eq:Zsde}
  \partial_x Z(x) = - \varepsilon\,\un - Z(x)^2 + V(x)
  \hspace{0.5cm}\mbox{with }
  \varepsilon = k^2
  \:,
\end{equation}
with the initial condition $Z(0) = \infty\,\un$, corresponding to the Dirichlet condition $\Psi(0)=0\,\un$. 
Eq.~\eref{eq:AsymptoticPhi} makes clear that the scattering matrix can be expressed as 
\begin{equation}
  \label{eq:LinkZS}
  \Sm = [k\,\un - \I Z(L)]\,[k\,\un+\I Z(L)]^{-1}
  \:,
\end{equation}
or equivalently
\begin{equation}
  Z(L) = \I k (\Sm-\un)(\Sm+\un)^{-1}
  \:.
\end{equation}
Using~(\ref{eq:Zsde}), we can write an equation describing the evolution of $\Sm$ upon increasing $L$:
\begin{equation}
  \label{eq:SDEforS}
    \partial_L \Sm 
    = 2 \I k\, \Sm  + \frac{1}{2\I k} (\un + \Sm) V(L) (\un + \Sm)
\end{equation}
One can check that this equation preserves the unitarity $\Sm^\dagger = \Sm^{-1}$. 
Additionally, for $\beta=1$, we have $V(x)^\T = V(x)$ therefore $\Sm^\T = \Sm$. 

Derivation of \eref{eq:SDEforS} with respect to $\varepsilon=k^2$ provides the MSDE satisfied by~$\WSm$~:
\begin{equation}
  \label{eq:SDEforQ}
  \hspace{-2.5cm}
    \partial_L \WSm = \frac{\un}{k}
    + \frac{1}{2\I k} \left[ \WSm\, V(L)\, (\un+\Sm) - (\un + \Sm^\dagger)\,V(L)\,\WSm
    \right]
    +\frac{1}{4k^3} (\un+\Sm^\dagger) V(L) (\un + \Sm)
    \:.
\end{equation}
In the next section, we analyse these equations in the weak disorder limit and identify fast and slow variables. 
Elimination of fast variables leads to simplified MSDE describing the variables on large scales.


\section{Averaging over fast variables in the weak disorder limit}
\label{sec:WeakDisEff}

In the weak disorder limit $\sigma \ll k^3$, the evolution of $\Sm$ and $\WSm$ is controlled by two length scales~:
\begin{itemize}
\item 
  the wavelength $\lambdabar=1/k$ controls the fast oscillations (which are present in the absence of disorder, $V(x)=0$);
\item 
  the localisation length $\xi=8k^2/ \sigma\gg\lambdabar$, or the mean free path $\ell_e\sim\xi/\Nc$, which is the typical length scale for the evolution of the other variables.
\end{itemize}
The idea is to perform some averaging over short scale $\lambdabar=1/k$ to get rid of the fast oscillations and obtain equations describing the evolution of $\Sm$ and $\WSm$ on the larger scale $\xi\sim (k^3/\sigma)\,\lambdabar$. 
The main difficulty is that MSDE, as Eq.~\eref{eq:SDEforS}, must be manipulated with care. 
A rigorous approach is to relate the MSDE to a Fokker-Planck equation for a matrix distribution~: 
this can be achieved for matrix random process \cite{Gra18}, however it is quite cumbersome. 
In the present paper, we discuss this approach in~\ref{sec:specCase} for the specific case $N=2$ and $\beta=1$. 
Here we have found more convenient to  work directly with MSDE by identifying effective independent noises.
We have kept some control on the method by comparing the outcome with the more rigorous Fokker-Planck approach in a specific case (\ref{sec:specCase}).

\subsection{The scattering matrix}

The starting point is to remove the fast oscillations by introducing
\begin{equation}
  \Smt = \EXP{-2\I k x}\, \Sm
\end{equation}
(from now on, $x$ must be understood as the size of the disordered region).
From Eq.~\eref{eq:SDEforS}, we obtain the MSDE satisfied by~$\Smt$~:
\begin{equation}
  \partial_x \Smt =
  \frac{1}{2\I k} (\EXP{-\I k x} + \Smt \EXP{\I k x}) \,
  V(x) \,
  (\EXP{-\I k x} + \EXP{\I k x} \Smt)
  \:.
\end{equation}
Thus
\begin{eqnarray}
  \partial_x \Smt = \frac{1}{2\I k}
  &
  \left[
    (\un + \Smt)\, \cos kx - \I\, (\un - \Smt)\, \sin kx
  \right]
  V(x)
  \nonumber\\
  &
  \hspace{0.5cm}
  \times
  \left[
    \cos kx\, (\un + \Smt) - \I \sin kx\, (\un - \Smt)
  \right]
  \:.
\end{eqnarray}
We can rewrite this equation as
\begin{equation}
  \label{eq:SDE_S_tilde}
  \hspace{-1cm}
    \partial_x \Smt =
    \frac{1}{2\I k} \left\lbrace
      \left[V_1(x) - \I\, V_2(x)\right]
      + \Smt  \left[V_1(x) + \I\, V_2(x)\right] \Smt
      + \Smt\, V(x)
      + V(x)\, \Smt
    \right\rbrace
    \:,
\end{equation}
where we have introduced
\begin{equation}
  V_1(x) = \cos (2kx) \, V(x)
  \hspace{0.5cm}\mbox{and}\hspace{0.5cm}
  V_2(x) = \sin (2kx) \, V(x)
  \:.
\end{equation}
In the weak disorder limit, the trigonometric functions oscillate fast compared to the typical length scale for the evolution of $\Smt$. 
In this limit, $V_1$, $V_2$ and $V$ become independent Gaussian white noises, as we now demonstrate.
Let us compute the correlations between the different processes~:
\begin{eqnarray}
  \hspace{-2cm}
  \mean{\int_0^x (V_1)_{ab} \int_0^{x'} (V_1)_{cd}^*}
  = \sigma \int_0^{\min{x}{x'}} C_{ab,cd}\, \cos^2(2kt)\, \D t
  \simeq \frac{\sigma}{2} C_{ab,cd}\, \min{x}{x'}
  \:,
\\
  \hspace{-2cm}
  \mean{\int_0^x (V_1)_{ab} \int_0^{x'} (V_2)_{cd}^*}
  = \sigma \int_0^{\min{x}{x'}} C_{ab,cd}\, \cos(2kt) \sin(2kx)\, \D t
  \simeq 0
  \:,
\\
  \hspace{-2cm}
  \mean{\int_0^x (V_1)_{ab} \int_0^{x'} V_{cd}^*}
  = \sigma \int_0^{\min{x}{x'}} C_{ab,cd}\, \cos(2kt)\, \D t
  \simeq 0
  \:.
\end{eqnarray}
The same properties holds for $V_2$. 
This shows that $V_1$, $V_2$ and $V$ become three independent Gaussian white noises, with
\begin{equation}
  V_1(x) \eqlaw 
  V_2(x) \eqlaw \frac{1}{\sqrt{2}} V(x)
  \:.
\end{equation}

\paragraph{Remark~:} 
from our derivation of~\ref{sec:specCase}, Eq.~\eref{eq:SDE_S_tilde} must be interpreted in the Stratonovich
sense.


\subsection{The Wigner-Smith matrix }

We introduce $\Sm = \EXP{2\I kx}\, \Smt$ in the MSDE \eref{eq:SDEforQ}~:
\begin{eqnarray}
  \nonumber
  \partial_x \WSm
  =
     \frac{\un}{k} 
     &+
     \frac{1}{2\I k}
     \left\lbrace
     \WSm \left[
     V + V_1 \Smt + \I V_2 \Smt
     \right]
     - \left[
     V + \Smt^\dagger V_1 - \I \Smt^\dagger V_2
     \right] \WSm
     \right\rbrace
  \\
   &+ \frac{1}{4k^3}
     \left[
     V + \Smt^\dagger V \Smt + \Smt^\dagger (V_1-\I V_2)
     + (V_1 + \I V_2) \Smt
     \right]
     \:.
     \label{eq:SDEforQinterm}
\end{eqnarray}
In the high energy limit, we can drop the last term of~\eref{eq:SDEforQinterm} which is subleading (anticipating on the result, $\WSm$ typically grows exponentially with the system size, while the neglected term is bounded).
We obtain
\begin{equation}
  \label{eq:SDEQlargeK}
    \partial_x \WSm \simeq \frac{\un}{k} +
    \frac{1}{2\I k}
    \left\lbrace
      \WSm \left[
        V + V_1 \Smt + \I V_2 \Smt
      \right]
      - \left[
        V + \Smt^\dagger V_1 - \I \Smt^\dagger V_2
      \right] \WSm
    \right\rbrace
    \:.
\end{equation}
We recall that $\Smt$ satisfies~(\ref{eq:SDE_S_tilde}).
Both~\eref{eq:SDE_S_tilde} and~\eref{eq:SDEQlargeK} must be interpreted in the Stratonovich sense.


\section{The \og square-root trick \fg{} and a new symmetrisation of the Wigner-Smith matrix }
\label{sec:NewSym}

In chaotic cavities, an important step for the determination of the distribution of the Wigner-Smith matrix eigenvalues was the introduction of the symmetrised Wigner-Smith matrix $\WSm_s=\Sm^{1/2}\WSm\Sm^{-1/2}$~\cite{BroFraBee99}.
This makes $\Sm$ and $\WSm_s$ independent and ensures that $\WSm_s$ is real symmetric for $\beta=1$. 
However, such a symmetrisation is not possible for multichannel 1D wires as we cannot get a simple MSDE satisfied by $\Sm^{1/2}$. 
To circumvent this problem we have to follow here a different strategy~: we introduce two unitary matrices $\UL$ and $\UR$ which satisfy the
equations
\begin{eqnarray}
  \label{eq:SDEforUL}
  \partial_x \UL
  &=
    \frac{1}{2\I k}
    \left(
    \frac{1}{2} (V_1 - \I V_2)\, \UR^{-1}
    + \frac{1}{2} \UL\, \UR \,(V_1+\I V_2)\, \UL
    + V \, \UL
    \right)
    \:,
  \\
  \label{eq:SDEforUR}
  \partial_x \UR
  &=
    \frac{1}{2\I k}
    \left(
    \frac{1}{2} \UL^{-1}\, (V_1 - \I V_2)
    + \frac{1}{2} \UR\, (V_1+\I V_2)\, \UL\, \UR
    + \UR \, V
    \right)
    \:.
\end{eqnarray}
One can easily check that these equations preserve the unitarity of both $\UL$ and $\UR$. 
Furthermore, we can deduce from (\ref{eq:SDEforUL},\ref{eq:SDEforUR}) a SDE for the matrix $\UL \UR$, which coincides with Eq.~\eref{eq:SDE_S_tilde}, thus
\begin{equation}
  \boxed{ 
  \Smt = \UL \UR 
  }
\end{equation}
This provides a factorisation of the scattering matrix which can be used to take some sort of ``\textit{square root}'' (a similar trick was used in \cite{MarSchBee16} in the orthogonal case). 
Furthermore, for orthogonal symmetry class, we can easily check that $\UR = \UL^\T$, thus 
\begin{equation}
  \Smt = \UL \UL^\T=\UR^\T\UR=\Smt^\T
  \hspace{0.5cm} \mbox{for }\beta=1
  \:.
\end{equation}

This allows us to introduce an alternative symmetrisation of the Wigner-Smith matrix 
\begin{equation}
  \boxed{
  \WSmt = \UR\, \WSm \, \UR^\dagger
  }
  = \EXP{-2 \I k x} \UL^\dagger \: \partial_\varepsilon (\EXP{2\I k x} \UL \UR) \:
  \UR^\dagger
  \:,
\end{equation}
where we have used that $\Sm = \EXP{2\I k x} \Smt = \EXP{2\I k x} \UL \UR$.

We can obtain the MSDE satisfied by $\WSmt$ by combining Eqs.~(\ref{eq:SDEQlargeK},\ref{eq:SDEforUR}). We thus obtain
\begin{equation}
  \label{eq:SDEforQs}
    \partial_x \WSmt = \frac{\un}{k}
    + \frac{1}{2 k}
    \left(
      \WSmt \,  W[\UL,\UR,V_1,V_2]
      + W[\UL,\UR,V_1,V_2] \, \WSmt
    \right)
    \:,
\end{equation}
where we have introduced the Hermitian matrix
\begin{equation}
  \label{eq:defW}
  \hspace{-1cm}
  W[\UL,\UR,V_1,V_2]
  =
  \frac{1}{2\I} \left( \UR \, V_1 \, \UL - \UL^\dagger \, V_1  \, \UR^\dagger \right)
  + \frac{1}{2}
  \left( \UR \,  V_2  \, \UL + \UL^\dagger \, V_2  \, \UR^\dagger  \right)
  \:.
\end{equation}
Integrating~\eref{eq:SDEforQs} over $[0,L]$ leads to 
\begin{equation}
  \label{eq:exprQsFctX}
    \WSmt = \frac{1}{k} \:
    X(L)
    \left(
      \int_0^L \D x \, X(x)^{-1} X^{\dagger}(x)^{-1} 
    \right)
    X^\dagger(L)
    \:,
\end{equation}
where $X(x)$ solves the MSDE
\begin{equation}
  \label{eq:EvolX}
  \partial_x X(x) = \frac{1}{2k} W[\UL,\UR,V_1,V_2] \: X(x)
  \:.
\end{equation}
The problem is now to study the equation~(\ref{eq:EvolX}), with $\UL$
and $\UR$ which satisfy respectively~(\ref{eq:SDEforUL})
and~(\ref{eq:SDEforUR}).

We stress that, up to now, we have made no assumption on the distribution of the random potential $V$ (and $V_1\eqlaw V_2 \eqlaw V/\sqrt{2}$), except that it is Gaussian.


\section{Isotropic case~: decoupling of $\Smt$ and $\WSmt$}
\label{sec:IsoCase}

We now rescale the matrix Gaussian white noise as $V\to\sqrt{\sigma}\,\eta$, with $\mean{\eta_{ab}(x)\eta_{cd}^*(x')}=C_{ab,cd}\,\delta(x-x')$.  
In this section, we use the mathematical notation for SDE, based on 
$\D B(x) = \eta(x)\,\D x$ satisfying
\begin{equation}
  \label{eq:CorrForB}
  \hspace{-1cm}
  \D B_{ab}(x)\D B_{cd}^*(x) = C_{ab,cd}\, \D x
  \hspace{0.5cm}\mbox{with }
    C_{ab,cd}
  = \frac{\beta}{2}\,\delta_{ac} \delta_{bd}+\left(1-\frac{\beta}{2}\right)\,\delta_{ad} \delta_{bc}
\end{equation}
We deduce the useful relation 
\begin{equation}
    \label{eq:NoiseSquareSandwich}
   \D B(x)\, \mathcal{O} \, \D B(x)
  =
  \left[
  \frac{\beta}{2}\, \tr{\mathcal{O}} \, \un
  + \left(1- \frac{\beta}{2}\right)\,\mathcal{O}^\T
  \right]\D x
\end{equation}
for any matrix $\mathcal{O}$ uncorrelated with $\D B(x)$.
In particular, setting $\mathcal{O}=\un$, we get 
\begin{equation}
      \label{eq:NoiseSquare}
  \D B(x)^2 = \mu\, \D x\, \un 
  \hspace{0.5cm}\mbox{where } 
  \mu  =1 + \frac{\beta}{2}(\Nc-1)
  \:.
\end{equation}

\subsection{Warm up~: case $\Nc=1$}
\label{subsec:WarmUp}

It is helpful to start the analysis by considering the case $\Nc=1$~:
averaging over the fast variable was performed in the Fokker-Planck equation in Ref.~\cite{ComTex97} (see also \cite{FarTsa94}).
Let us see how Eqs.~\eref{eq:SDEQlargeK} and~\eref{eq:SDE_S_tilde} yield the known result \eref{eq:WTDfBM} by manipulating the SDE. 
Let us denote $\Smt = \EXP{\I\alpha}$. 
Eqs.~(\ref{eq:SDEQlargeK},\ref{eq:SDE_S_tilde}) reduce to
\begin{eqnarray}
  \D \alpha(x)
  &=
  -\frac{\sqrt{\sigma}}{k} 
    \left[ 
      \D B(x) 
      + \frac{1}{\sqrt{2}}
      \left(
       \cos \alpha \, \D B_1(x)  - \sin \alpha \, \D B_2(x) 
     \right)     \right]
    \:,
  \\
  \D \WSm(x)
  &=
    \frac{1}{k} 
    \left[
      \D x + 
      \frac{\sqrt{\sigma}}{\sqrt{2}}
      \left(
         \sin \alpha \, \D B_1(x) 
       + \cos \alpha\,  \D B_2(x) 
      \right) \WSm
    \right]
    \:,
\end{eqnarray}
where $B(x)$, $B_1(x)$ and $B_2(x)$ are three independent normalised Brownian motions.
As mentioned above, these two equations are interpreted in the Stratonovich sense.
Relating them to SDE in the It\^o sense, we get here the same equations.
Let us now choose the It\^o convention for convenience.
We define two new noises
\begin{eqnarray}
  \D w_1(x) &= \cos \alpha(x) \: \D B_1(x)  - \sin \alpha(x) \: \D B_2(x) 
  \\
  \D w_2(x) &= \sin \alpha(x) \:  \D B_1(x)  + \cos \alpha(x) \: \D B_2(x) 
  \:.
\end{eqnarray}
Since we work with the It\^o convention, we have $\mean{\D w_1(x)} = \mean{\D w_2(x)} = 0$.  The strength
of the noises is
$
  \D w_1(x)^2  = \D w_2(x)^2 = \D x
$
and they are clearly uncorrelated, $\D w_1(x) \D w_2(x) = 0$.
The two new noises are thus independent, and we can rewrite
\begin{eqnarray}
  \D \alpha(x)
  =
  -\frac{\sqrt{\sigma}}{k}  \left[ \D B(x) + \frac{1}{\sqrt{2}}\D w_1(x)  \right]
  \hspace{1cm}
  \mbox{ (It\^o)}
  \\
  \D \WSm(x)
  =
  \frac{1}{k}\left[
    \D x + \frac{\sqrt{\sigma}}{\sqrt{2}}\,\D w_2(x)\, \WSm
  \right]
  \hspace{1.75cm}
  \mbox{ (It\^o)}
\end{eqnarray}
All the manipulations have assumed that SDE are in the It\^o sense. 
Converting the second equation to Stratonovich convention, we obtain
\begin{equation}
  \partial_x \WSm \eqlaw \frac{1}{k} +
  \left( \frac{V(x)}{ \sqrt{2} k} -\frac{\sigma}{4k^2}  \right) \WSm
  =\frac{1}{k} + \left( \frac{2\,\eta(x)}{\sqrt{\xi}} - \frac{2}{\xi} \right) \WSm
  \:,
\end{equation}
where $V(x)$ is the original potential and $\eta(x)$ a normalised Gaussian white noise. 
Thus, we have recovered the result of Ref.~\cite{ComTex97} and Eq.~\eref{eq:WTDfBM}, following a more simple procedure.

\subsection{Strategy for $\Nc>1$}

Let us now consider the case of isotropic noise, which corresponds to a correlator of the form~\eref{eq:Ciso}. We consider Eq.~\eref{eq:EvolX} instead of the symmetrised Wigner-Smith matrix $\WSmt$, since they can be easily related via~\eref{eq:exprQsFctX}.

Let us first rewrite
Eqs.~(\ref{eq:EvolX},\ref{eq:SDEforUL},\ref{eq:SDEforUR}) in the form
\begin{eqnarray}
  \label{eq:SDEforX}
  \D X
  &=
    \frac{\sqrt{\sigma/2}}{2k} \, W[\UL,\UR,\D B_1, \D B_2] \, X
  \\
  \label{eq:SDEforULb}
  \D \UL
  &=
    \frac{\sqrt{\sigma}}{2\I k}
    \left(
    \frac{1}{\sqrt{2}} \, \UL \, W_u[\UL,\UR,\D B_1,\D B_2]
    + \D B(x) \, \UL
    \right)
    \:,
  \\
  \label{eq:SDEforURb}
  \D \UR
  &=
    \frac{\sqrt{\sigma}}{2\I k}
    \left(
    \frac{1}{\sqrt{2}} \, W_u[\UL,\UR,\D B_1,\D B_2] \,  \UR
    + \UR\,  \D B(x)
    \right)
    \:,
\end{eqnarray}
where $W$ is given by Eq.~(\ref{eq:defW}) and we have denoted
\begin{equation}
  \label{eq:Wu}
  \hspace{-2cm}
  W_u[\UL,\UR,\D B_1, \D B_2] =
  \frac{1}{2}
  \left(
    \UL^\dagger \, \D B_1 \, \UR^\dagger + \UR \, \D B_1 \, \UL
  \right)
  + \frac{1}{2 \I}
  \left(
    \UL^\dagger\, \D B_2 \, \UR^\dagger - \UR \, \D B_2 \, \UL
  \right)
  \:.
\end{equation}
$B_1$, $B_2$ and $B$ are now three independent normalised Brownian motions, each satisfying~\eref{eq:CorrForB}.  The idea is the following~: since the $B_i$'s are isotropic, the noises $W$ and $W_u$ can be shown to be independent, and we can thus decouple the equations for $\UL$ and $\UR$ from the equation on $X$ (and thus $\WSmt$).

In order to do so, the procedure is the following:
\begin{enumerate}
\item 
  Convert the stochastic equations from Stratonovich to It\^o convention in order to decouple the matrices from the noises at coinciding points;
\item 
  Show that the two noises $W$ and $W_u$ are independent Gaussian white noises (independently of $\UL$ and $\UR$), and then replace them with new ones with the same distribution, but which do no involve $\UL$ or $\UR$;
\item 
  Convert the new equations back to Stratonovich convention.
\end{enumerate}
Concerning the first point, we only need to convert the equation on
$X$, since we will no longer be interested in the unitary
matrices.

\subsection{Conversion to the It\^o convention}

Converting a stratonovich MSDE \eref{eq:SDEforX} into the It\^o convention brings an additional drift~ \footnote{
  \label{footnote:StratoIto}
  The simplest way to perform the Stratonovich$\to$It\^o conversion is as follows.
  Consider the Stratonovich SDE
  $\D x(t) = \alpha(x(t))\, \D t + b(x(t))\,\D W(t)$, 
  and define $(\D x)_\mathrm{noise} = b(x)\,\D W(t)$.
  The corresponding It\^o equation is obtained by writing
  $
  \D x(t) = \alpha(x)\, \D t + b\!\left( x + \frac{1}{2} (\D x)_\mathrm{noise}  \right)\,\D W(t)
  = a(x)\, \D t + b(x)\,\D W(t)
  $ 
  with 
  $
  a(x)=  \alpha(x) +  \frac{1}{2} b'(x) b(x)  
  $, 
  where we have used $\D W(t)^2=\D t$.
  This simple procedure can be applied whatever the nature of the process is (scalar, vector, matrix,...).
}
\begin{equation}
  \ItoDr =
  \frac{1}{2} \sum_{i,j} \left(
  (\D X)_{ij}  \derivp{(\D X)}{X_{ij}}
  + (\D \UL)_{ij} \derivp{(\D X)}{(\UL)_{ij}}
  + (\D \UR)_{ij} \derivp{(\D X)}{(\UR)_{ij}}
  \right)
  \:.
\end{equation}
The MSDE $\D X$~(\ref{eq:SDEforX}) depends on $\UL$ and $\UR$ only through $W$, which is linear in $\UL$, $\UR$, $\UL^\dagger$ and $\UR^\dagger$. Thus,
\begin{eqnarray}
  \nonumber
  \ItoDr
  &=
     \frac{\sqrt{\sigma/2}}{4k}
     \left( W[\UL,\UR,\D B_1,\D B_2] \right.
     \D X
  \\
  \nonumber
   & \hspace{2cm} + \left. W[\D \UL,\UR,\D B_1,\D B_2]\, X
     + W[\UL,\D \UR,\D B_1,\D B_2]\,  X
     \right)
  \\
  &=
     \frac{\sqrt{\sigma/2}}{4k}
     \Bigg(
     \frac{\sqrt{\sigma/2}}{2k} W[\UL,\UR,\D B_1,\D B_2]^2
  \\
  \nonumber
   &\hspace{2cm}+ W[\D \UL,\UR,\D B_1,\D B_2]
     + W[\UL,\D \UR,\D B_1,\D B_2]
     \Bigg) X
     \:.
\end{eqnarray}
Let us look at the second term. When replacing $W$ and $\D \UL$ by their expressions, we obtain products of the different noises $\D B$, $\D B_1$ and $\D B_2$. Since they are independent, the only non vanishing terms will involve products of the same noise. We thus obtain
\begin{eqnarray}
  \hspace{-1cm}
  \nonumber
  W[\D \UL,\UR, \D B_1,\D B_2]
  =
  &- \frac{\sqrt{\sigma/2}}{4k}
  \Bigg(
  \UR \, \D B_1^2 \, \UR^\dagger
  + \UR \, \D B_1 \, \UL \UR \, \D B_1 \, \UL
  \\
  & + \UR \, \D B_2^2 \,  \UR^\dagger - \UR \, \D B_2 \, \UL \UR \, \D
    B_2 \, \UL
  + \text{h.c.}
  \Bigg)
  \:.
\end{eqnarray}
Since $\D B_1(x) \eqlaw \D B_2(x) \eqlaw \D B(x)$, this reduces to
\begin{equation}
  \label{eq:ConvItoUL}
  W[\D \UL,\UR,V_1,V_2] =
  - \frac{\sqrt{\sigma/2}}{2k} \, \UR \, \D B^2 \, \UR^\dagger
  \:,
\end{equation}
Similarly,
\begin{equation}
  \label{eq:ConvItoUR}
  W[\UL,\D \UR,V_1,V_2] =
  - \frac{\sqrt{\sigma/2}}{2k} \, \UL^\dagger \, \D B^2 \, \UL
  \:.
\end{equation}
So that we finally get
\begin{equation}
  \label{eq:ConvItoAll}
  \hspace{-1cm}
  \ItoDr
  = \frac{\sigma}{16 k^2} W[\UL,\UR,\D B_1,\D B_2]^2 X
  - \frac{\sigma}{16 k^2} \left(
    \UR \, \D B^2 \, \UR^\dagger
    + \UL^\dagger \, \D B^2 \, \UL
  \right) X
  \:.
\end{equation}
This equation holds for any type of correlated matrix noise $\D B_1(x)$.  
Now making the assumption that the noise is isotropic, we deduce from~\eref{eq:NoiseSquare},
\begin{equation}
  \label{eq:ItoDrift}
   \ItoDr
  = \frac{\sigma}{16 k^2} W[\UL,\UR,\D B_1,\D B_2]^2 X
  - \frac{\sigma}{8k^2}\left( 1 + \beta \frac{N-1}{2}  \right) X \D x
  \:.
\end{equation}
We do not evaluate the first term now, as it will cancel out in the following when we will get back to the Stratonovich form. 
Nevertheless, it can be easily evaluated from the correlator of $W$, which we now analyze.

\subsection{Characterisation of the effective noises $W$ and $W_u$}

Let us now study the distribution of the noises
$W[\UL,\UR,\D B_1,\D B_2]$ and $W_u[\UL,\UR,\D B_1,\D B_2]$, given
respectively by Eqs.~\eref{eq:defW} and~\eref{eq:Wu}. We start by
computing
\begin{eqnarray}
  \nonumber
  \hspace{-2cm}
  W_{ab}W^*_{cd} = W_{ab} W_{dc}
  = -\frac{1}{4}
    \left( \UR \, \D B_1 \, \UL - \UL^\dagger \, \D B_1 \, \UR^\dagger \right)_{ab}
    \left( \UR \, \D B_1 \, \UL - \UL^\dagger \, \D B_1 \, \UR^\dagger \right)_{dc}
  \\
  + \frac{1}{4} 
    \left( \UR \, \D B_2 \, \UL + \UL^\dagger \, \D B_2 \, \UR^\dagger  \right)_{ab}
    \left( \UR\,  \D B_2 \, \UL + \UL^\dagger \, \D B_2 \, \UR^\dagger  \right)_{dc}
\end{eqnarray}
Expanding and keeping only the non-vanishing terms, and using that
$\D B_1 \eqlaw \D B_2 \eqlaw \D B$, we obtain
\begin{eqnarray}
  \label{eq:CorWGen}
  \hspace{-2cm}
  W_{ab}W^*_{cd}
  \\\nonumber
  \hspace{-1.5cm}
  = \frac{1}{2} \sum_{pqrs}
  \left(
    (\UR)_{ap} (\UL)_{qb} (\UL^\dagger)_{dr} (\UR^\dagger)_{sc}
    + (\UL^\dagger)_{ap} (\UR^\dagger)_{qb} (\UR)_{dr} (\UL)_{sc}
  \right)
  \D B_{pq}(x) \D B_{sr}(x)^*
  \:.
\end{eqnarray}
In the isotropic case, using the expression of the
correlator~(\ref{eq:CorrForB}), we get
\begin{equation}
  W_{ab}[\UL,\UR,\D B_1,\D B_2] \, W^*_{cd}[\UL,\UR,\D B_1,\D B_2]
  = C_{ab,cd}\, \D x
  \:.
\end{equation}
Similarly, we obtain
\begin{eqnarray}
  \hspace{-2cm}
  W_{ab}(W_u)^*_{cd} =
  \\\nonumber
  \hspace{-1.5cm}
  \frac{1}{4\I} \sum_{pqrs}
  \left(
    (\UR)_{ap} (\UL)_{qb} (\UL^\dagger)_{dr} (\UR^\dagger)_{sc}
    - (\UL^\dagger)_{ap} (\UR^\dagger)_{qb} (\UR)_{dr} (\UL)_{sc}
  \right)
  \D B_1(x)_{pq} \D B_1(x)^*_{sr}
  \:.
\end{eqnarray}
Which, in the isotropic case yields
\begin{equation}
  W_{ab}[\UL,\UR,\D B_1,\D B_2] (W_u^*)_{cd}[\UL,\UR,\D B_1,\D B_2] = 0
  \:,
\end{equation}
showing that $W$ and $W_u$ are uncorrelated. Therefore, we have shown
that 
\begin{equation}
  W[\UL,\UR,\D B_1(x), \D B_2(x)] \eqlaw \D B(x)
 \:,
\end{equation}
so that we can
rewrite the MSDE~\eref{eq:SDEforX} as
\begin{equation}
  \label{eq:EvolXIto}
  \D X = \big(\ItoDr\big)\: X + \frac{\sqrt{\sigma/2}}{2 k} \D B(x)\, X
  \hspace{0.5cm} \text{(It\^o),}  
\end{equation}
where the drift is given by Eq.~\eref{eq:ItoDrift}. 
Isotropy has been used to perform unitary transformations such that the unitary matrices $\UL$ and $\UR$ can be removed from the MSDE for $X(x)$. 
Eventually, we have obtained a MSDE which involves no other matrix than~$X$.

\subsection{Back to the Stratonovich convention}

We can now convert back the It\^o equation \eref{eq:EvolXIto} into a Stratonovich one. 
One has to add the drift term
\begin{equation}
 - \frac{1}{2} \sum_{ij}
   (\D X)_{ij} \derivp{(\D X)}{X_{ij}}
  = -\frac{\sigma}{16 k^2} \D B(x)^2 X
\end{equation}
 to the It\^o equation.
This cancels out the first term in~\eref{eq:ItoDrift}, and we thus get
\begin{equation}
  \label{eq:EvolXStrato}
    \partial_x X =
    - \frac{\mu}{\xi}\, X  + \frac{1}{\sqrt{\xi}} \eta(x)\, X
  \hspace{0.5cm} \text{(Stratonovich)}  
\end{equation}
where we recall that $\mu=1+ \beta \frac{N-1}{2}$ and $\D B(x) = \eta(x) \D x$, so that  $\eta(x)$ is a Hermitian Gaussian white noise.  
Therefore, the matrix $X$ is an exponential of a matrix Brownian motion, Eq.~\eref{eq:LambdaChronological}.  
The symmetrised Wigner-Smith matrix $\WSmt$ is expressed as a functional of this exponential of Brownian motion via Eq.~\eref{eq:exprQsFctX}.  
This extends the result known for $N=1$ to higher number of channels.

From the expression~\eref{eq:exprQsFctX} of $\WSmt$ and the stochastic equation on $X$~\eref{eq:EvolXStrato}, we can derive the MSDE
\begin{equation}
  \label{eq:MSDEforQt}
  \boxed{
  \derivp{}{L} \WSmt = \frac{\un}{k}
  - \frac{2 \mu}{\xi} \WSmt
  + \frac{1}{\sqrt{\xi}} \left( \WSmt\, \eta(L) + \eta(L) \, \WSmt \right)
  \hspace{0.5cm} \text{(Stratonovich)}  
  }
\end{equation}
This equation is a central result that will be used below.

%


\section{Matrix generalization of the Dufresne identity}
\label{sec:MatrixDufresne}

In this Section, we discuss the relation with the work of Rider and Valk\'o~\cite{RidVal15} and extend their result.
We have obtained the symmetrised Wigner-Smith matrix under the form of an exponential functional of the matrix BM \eref{eq:exprQsFctX}.
A first difference with Rider and Valk\'o's functional concerns the form of the integral. 
A second difference is that the matrix BM of Rider and Valk\'o involves a non Hermitian noise with $\Nc^2$ independent real entries (orthogonal class), while we have considered a Hermitian real or complex noise (orthogonal or unitary class).

\subsection{Exponential functional}
\label{subsubsec:EF}

Let us introduce 
\begin{equation}
  \Lambda(x/\xi)^\dagger = X(L) X(L-x)^{-1}
  \:,
  \quad
  x \in [0,L]
  \:.
\end{equation}
From~\eref{eq:EvolXStrato}, we get that $\Lambda$ satisfies the MSDE
\begin{equation}
  \label{eq:MSDEforLambda}
  \partial_x \Lambda =   - \mu\, \Lambda  +  \eta(x) \, \Lambda
  \:,
\end{equation}
where $\eta(x)$ is the Hermitian Gaussian white noise and $\mu=1+\beta(\Nc-1)/2$.
The initial condition is obviously $\Lambda(0)=\un$.
The representation \eref{eq:exprQsFctX} can be rewritten in a more simple form with the new matricial random process~:
\begin{equation}
  \label{eq:QtildeForRidVal}
  \boxed{
  \WSmt
  \eqlaw 2\tau_\xi 
  \int_0^{L/\xi}\D x\,  \Lambda(x)^\dagger \Lambda(x) 
  }
\end{equation}

\subsection{\og Gauge \fg{} transformation}

Rider and Valk\'o in Ref.~\cite{RidVal15} have considered the matricial stochastic process
\begin{equation}
  \label{eq:MSDE-RiderValko}  
  \partial_x M(x) = -\mu\,M(x) + \chi(x)\,M(x)
  \:,
\end{equation}
where $\chi(x)$ is a $\Nc\times\Nc$ real matrix, whose elements are $\Nc^2$ independent normalised \textit{real} Gaussian white noises. Thus the noise matrix is \textit{non Hermitian}, $\chi(x)\neq\chi(x)^\dagger$.
Let us rather consider \eref{eq:MSDE-RiderValko}  when the matrix elements of $\chi(x)$ are $\Nc^2$ independent \textit{complex} noises.
We decompose it into Hermitian and anti-Hermitian parts:
\begin{equation}
  \chi = \eta + \I\, A
  \:,
  \qquad
  \eta = \frac{\chi + \chi^\dagger}{2}
  \:,
  \quad
  A = \frac{\chi-\chi^\dagger}{2\I}
  \:,
\end{equation}
thus $\mean{\eta_{ab}(x)\eta_{cd}(x')}=C_{ab,cd}\,\delta(x-x')$.
In order to relate \eref{eq:MSDE-RiderValko} to \eref{eq:MSDEforLambda}, we \og gauge out \fg{}  the non-Hermitian part 
\begin{equation}
  \label{eq:GaugeTransf}
  M = U \, \Lambda
  \hspace{1cm}\mbox{where}\hspace{1cm}
  \partial_x U = \I\, A\, U
  \:.
\end{equation}
It is straightforward to get 
\begin{equation}
  \partial_x \Lambda(x) =   - \mu\, \Lambda(x)  + U(x)^{-1}\, \eta(x)\, U(x) \, \Lambda(x)
  \:.
\end{equation}
These equations are understood in the Stratonovich convention.
We use the mathematical notation 
$\D \Lambda =   - \mu\, \Lambda\, \D x  + U^{-1}\, \D B(x)\, U \, \Lambda$, where 
$\D B(x)=\eta(x)\,\D x$. 
Let us now go to the It\^o convention.
Using Eq.~\eref{eq:NoiseSquare}, we get 
\begin{eqnarray}
  \D \Lambda &=   - \frac{\mu}{2}\, \Lambda\, \D x  + U^{-1}\, \D B(x)\, U \, \Lambda
  \eqlaw  - \frac{\mu}{2}\,  \Lambda\, \D x  + \D B(x) \, \Lambda
  \hspace{0.5cm}\mbox{(It\^o)}
  \:.
\end{eqnarray}
Now going back to the Stratonovich convention, we get 
$\D \Lambda = - \mu\,\Lambda\, \D x  + \D B(x) \, \Lambda$, i.e. Eq.~\eref{eq:MSDEforLambda}.

\subsection{Matrix Dufresne identity}

Using the connection with the MSDE studied in Ref.~\cite{GraTex16}, we generalize in this paragraph Rider and Valk\'o's result for $\beta=1$ to both symmetry classes ($\beta=1$ and $2$).
The matricial process studied in Ref.~\cite{GraTex16}, which arises in a different multichannel localization model, is 
\begin{equation}
  \label{eq:MSDE-GT2016}
  \partial_x\Riccati 
  =|\varepsilon|^2 -2\mu\,g\,\Riccati - \Riccati^2 
  + \sqrt{g}\,\bi[\eta(x)\,\Riccati+\,\Riccati\,\eta(x)\big]
  \:.
\end{equation}
It was shown to be characterized by the stationary matrix distribution (in the $x\to\infty$ limit)
\begin{equation}
  \label{eq:GrabschTexier2016}
  P(\Riccati)
  \propto
  \left(\det\Riccati\right)^{-1-\mu-\beta(\Nc-1)/2}
  \exp\left[-\frac{1}{2g}\tr{\Riccati + |\varepsilon|^2\Riccati^{-1}} \right]
  \:.
\end{equation}
The distribution was obtained from the analysis of the related matrix Fokker-Planck equation (the case where the noise $\eta$ is not isotropic was also considered in \cite{GraTex16}).
The mapping to our problem can be easily realised~: we set
\begin{equation}
  \Riccati \to k\,|\varepsilon|^2\,\WSmt
  \hspace{0.5cm}\mbox{and}\hspace{0.5cm}
  g \to 1/\xi
\end{equation}
in Eqs.~(\ref{eq:MSDE-GT2016},\ref{eq:GrabschTexier2016}). 
For $ \mu>\beta(\Nc-1)/2$, it is possible to take the limit $|\varepsilon|\to0$ and recover the MSDE \eref{eq:MSDEforQt} while the distribution \eref{eq:GrabschTexier2016} takes the form
\begin{equation}
  P( \WSmt )    \propto
  \left(\det\WSmt\right)^{-1-\mu-\beta(\Nc-1)/2} \,\EXP{-\tau_\xi\tr{\WSmt^{-1}}} 
  \:.
\end{equation} 
The change of variable $\Gamma=2\tau_\xi\,\WSmt^{-1}$ leads to the Wishart distribution
\begin{equation}
  \label{eq:WishartForDufresne}
  \boxed{
  P(\Gamma) = C_{\Nc,\beta}\,
    (\det \Gamma)^{\mu-1-\beta(\Nc-1)/2}
  \EXP{-(1/2)\tr{\Gamma}}
  \hspace{0.5cm}\mbox{for } \mu>\frac{\beta(\Nc-1)}{2}
  }
\end{equation}
where $C_{\Nc,\beta}$ is a normalisation.
Having shown the relation between the MSDE \eref{eq:MSDEforQt} and the representation \eref{eq:QtildeForRidVal}, we conclude that 
\begin{equation}
 \left( \int_0^\infty \D x\, M(x)^\dagger M(x) \right)^{-1}
  = 
  \left( \int_0^\infty \D x\, \Lambda(x)^\dagger \Lambda(x) \right)^{-1}
   \eqlaw 
 \Gamma
  \:,
\end{equation}
 is distributed according to \eref{eq:WishartForDufresne}, where $\Lambda(x)$ solves \eref{eq:QtildeForRidVal} for a Hermitian noise and $M(x)$ solves \eref{eq:MSDE-RiderValko} for a non Hermitian noise. 
Here, the derivation was done for arbitrary drift~$\mu$, thus extending the Dufresne identity to the case of matrix BM, for orthogonal and unitary classes. 
Coming back to the multichannel disordered wire model, the drift is given by \eref{eq:DriftLoc}, leading to \eref{eq:LagWishDist}, i.e. to Beenakker \& Brouwer's result~\eref{eq:BeenakkerBrouwer2001}.


\section{Application~: moments $\mean{\tr{\WSm}}$, $\mean{\tr{\WSm}^2}$ and $\mean{\tr{\WSm^2}}$}
\label{sec:Moments}

In this section we show that the representation \eref{eq:QtildeForRidVal} in terms of exponential functional of the matrix BM, or equivalently the MSDE \eref{eq:MSDEforQt}, can be used to compute the moments of the form $\mean{\tr{\WSm^n}^m}$. We consider the first moments.
Our starting point is to rewrite the MSDE \eref{eq:MSDEforQt} in the It\^o convention.
We use the notation $\D B(x)=\eta(x)\,\D x$ as in the previous section.
Using (\ref{eq:NoiseSquareSandwich},\ref{eq:NoiseSquare}), we get~:
\begin{equation}
  \label{eq:MSDE-WSmt-Ito}
  \hspace{-2cm}
  \boxed{
  \D\WSmt = 
  \left( \frac{\un}{k} + \frac{\beta}{2\xi}\left[\tr{\WSmt} \un - \Nc\,\WSmt\right] \right)\D x
  +
  \frac{1}{\sqrt{\xi}}
  \left(
    \WSmt \, \D B(x) + \D B(x) \, \WSmt
  \right)
  \mbox{ (It\^o)}
  }
\end{equation}
We deduce immediately the equation for the trace 
\begin{equation}
  \D \tr{\WSmt}
  = \frac{\Nc}{k}\,\D x
  + \frac{2}{\sqrt{\xi} } \tr{\WSmt\, \D B(x)}  
  \hspace{0.5cm}\mbox{(It\^o)}
\end{equation}
and thus 
\begin{equation}
  \D \mean{\tr{\WSm}}
  = \frac{\Nc}{k}\,\D x
  \:,
\end{equation}
where we have used $\mathrm{tr}\big\{\WSmt^n\big\}=\tr{\WSm^n}$.
After integration over $[0,L]$, we recover the expected behaviour, Eq.~\eref{eq:MeanTraceQ},
\begin{equation}
   \mean{\tr{\WSm}} = \frac{\Nc L}{k} = \frac{2\Nc L}{v} 
   \:,
\end{equation}
where $v=2k$ is the group velocity.

Next, we write an equation for
$\big[\mathrm{tr}\big\{\WSmt\big\}\big]^2$.  Using
It\^o's formula~\footnote{ Using the notations of the
  footnote~\ref{footnote:StratoIto}, a simple way to recover
   It\^o's formula is as follwos~:
  $ \D f(x) = f(x+ \D x) - f(x) = f'(x) \, \D x + \frac12\, f''(x) \,
  \D x^2 = f'(x) \, \D x + \frac12\, f''(x) \, \big[(\D
  x)_\mathrm{noise}\big]^2 $.  This form is appropriate to be applied
  to the case of MSDE.  } we have
\begin{equation}
  \D\Big( \tr{\WSmt}^2 \Big)
  =  2 \tr{\WSmt}\,\D \tr{\WSmt} + \Big(\D \tr{\WSmt} \Big)^2
  \hspace{0.5cm}\mbox{(It\^o)}
\end{equation}
Using 
\begin{equation}
  \left( \tr{\WSmt\,\D B(x)} \right) ^2 = \tr{\WSmt^2}\D x
  \:,
\end{equation}
  we deduce 
\begin{equation}
  \label{eq:SDEtrace-sq}
  \hspace{-2cm}
  \D \tr{\WSmt}^2 
   = 
   \left[
     \frac{2\Nc}{k} \tr{\WSmt} + \frac{4}{\xi}\tr{\WSmt^2}
   \right]\D x
   + \frac{4}{\sqrt{\xi}}
   \tr{\WSmt}\,\tr{\WSmt\,\D B(x)}
  \mbox{ (It\^o)}
\end{equation}  
This makes clear that we have to derive also an equation for $\mathrm{tr}\big\{\WSmt^2\big\}$.
From \eref{eq:MSDE-WSmt-Ito}, the application of the It\^o formula gives 
\begin{eqnarray}
\hspace{-2cm}
  \D\left( \WSmt^2\right)
  = 
  \left\{
    \frac{2\WSmt}{k}
    + \frac{1}{\xi}
    \left[
      \frac{\beta}{2}\,\tr{\WSmt^2}\,\un  
      + 2\beta \,\tr{\WSmt}\,\WSmt
      + \left(
        2(2-\beta) - \frac{\beta\Nc}{2}
      \right) \WSmt^2
    \right]
  \right\}\D x
  \nonumber
  \\
  + \frac{1}{\sqrt{\xi} } 
  \left[\WSmt^2\, \D B(x) +  2\WSmt\, \D B(x)\,\WSmt +\D B(x)\,\WSmt^2  \right]
  \hspace{0.5cm}\mbox{(It\^o)}
\end{eqnarray}
A trace gives 
\begin{eqnarray}
  \label{eq:SDEsq-trace}
\hspace{-2cm}
  \D\tr{ \WSmt^2}
  = 
  \left( 
    \frac{2}{k}\tr{\WSmt}
    + \frac{4}{\xi}
    \left[
      \frac{\beta}{2}\,\tr{\WSmt}^2 
      + \left(1- \frac{\beta}{2}\right)  \tr{\WSmt^2} 
    \right]
  \right)\D x
  \nonumber
  \\
  + \frac{4}{\sqrt{\xi} } 
  \tr{ \WSmt^2\, \D B(x) }
  \hspace{0.5cm}\mbox{(It\^o)}
  \:.
\end{eqnarray}

Averaging (\ref{eq:SDEtrace-sq},\ref{eq:SDEsq-trace}), we deduce a simple linear problem
\begin{equation} 
  \label{eq:EqWhichisnomore114}
  \hspace{-1cm}
  \derivp{}{L}
  \left(\begin{array}{c}
    \mean{\tr{\WSm}^2} \\ \mean{\tr{\WSm^2}}
  \end{array}\right)
  = \frac{2NL}{k^2}
  \left(\begin{array}{c}
    \Nc \\ 1
  \end{array}\right)
  +\frac{4}{\xi}
  \left(
  \begin{array}{cc}
    0 & 1  \\ \frac{\beta}{2} & 1 - \frac{\beta}{2}
  \end{array}
  \right)
  \left(\begin{array}{c}
    \mean{\tr{\WSm}^2} \\ \mean{\tr{\WSm^2}}
  \end{array}\right)
\end{equation}
It is useful to use the spectral decomposition
\begin{equation}
  M_\beta = 
  \left(
  \begin{array}{cc}
    0 & 1  \\ \frac{\beta}{2} & 1 - \frac{\beta}{2}
  \end{array}
  \right)
  =  
  \sum_{\sigma=\pm} \lambda_\sigma\,\Pi_\sigma
  \:,
\end{equation}
where the eigenvalues $\lambda_\pm$ and the corresponding projectors are
\begin{eqnarray}
  \lambda_+ = 1 \hspace{0.5cm}\mbox{and}\hspace{0.5cm}
  \Pi_+ = 
  \frac{1}{1+\frac{\beta}{2}}
  \left(
  \begin{array}{cc}
    \frac{\beta}{2} & 1  \\ \frac{\beta}{2} & 1 
  \end{array}
  \right)
  \\
  \lambda_- = -\frac{\beta}{2} \hspace{0.5cm}\mbox{and}\hspace{0.5cm}
  \Pi_- = 
  \frac{1}{1+\frac{\beta}{2}}
  \left(
  \begin{array}{cc}
    1 & -1  \\ -\frac{\beta}{2} & \frac{\beta}{2}
  \end{array}
  \right)
  \:.
\end{eqnarray}
Integration of \eref{eq:EqWhichisnomore114} gives 
\begin{equation}
  \left(
  \begin{array}{c}
    \mean{\tr{\WSm}^2}
    \\
    \mean{\tr{\WSm^2}}
  \end{array}
  \right)   
  =\frac{2\Nc}{k^2}\int_0^L\D x\, x\, \EXP{\frac{4}{\xi}M_\beta\,(L-x)}
  \left(
  \begin{array}{c}
    \Nc
    \\
    1
  \end{array}
  \right)   
  \:.
\end{equation}
Using 
$\EXP{M_\beta\,y}=\sum_{\sigma} \EXP{\lambda_\sigma y}\,\Pi_\sigma$,
some simple algebra gives
\begin{eqnarray}
  \hspace{-1cm}
  \mean{\tr{\WSm}^2}
  =
  \frac{\Nc\tau_\xi^2}{2}
  \\\nonumber
  \times
  \left\{
    \frac{1 + \frac{\beta\Nc}{2}}{1+\frac{\beta}{2}}\,
    \left[\EXP{4L/\xi} - 1 - \frac{4L}{\xi}\right]
    +\frac{\big(\frac{2}{\beta}\big)^2(\Nc-1)}{1+\frac{\beta}{2}}\,
    \left[\EXP{-2\beta L/\xi} - 1 + \frac{2\beta L}{\xi}\right]
  \right\}
  \\
  \hspace{-1cm}
  \mean{\tr{\WSm^2}}
  =
  \frac{\Nc\tau_\xi^2}{2}
  \\\nonumber
  \times
  \left\{
    \frac{1 + \frac{\beta\Nc}{2}}{1+\frac{\beta}{2}}\,
    \left[\EXP{4L/\xi} - 1 - \frac{4L}{\xi}\right]
    -\frac{\frac{2}{\beta} (\Nc-1)}{1+\frac{\beta}{2}}\,
    \left[\EXP{-2\beta L/\xi} - 1 + \frac{2\beta L}{\xi}\right]
  \right\}
\end{eqnarray}
For $\Nc=1$, we recover 
$\mean{\WSm^2}=\frac{\tau_\xi^2}{2}\left(\EXP{4L/\xi} - 1 - \frac{4L}{\xi}\right)$
\cite{TexCom99}, as it should.

We remark that 
\begin{equation}
  \label{eq:RelationForCorrelationsBetweenProperTimes}
  \mean{\tr{\WSm}^2} - \mean{\tr{\WSm^2}}
  = \tau_\xi^2\,\Nc(\Nc-1)\,\frac{4}{\beta}\, 
  \left[\frac{L}{\xi}+\frac{\EXP{-2\beta L/\xi}-1}{2\beta}  \right]
  \:.
\end{equation}
Introducing the \textit{proper time delays} $\{\tau_a\}_{a=1,\cdots,\Nc}$, the eigenvalues of the Wigner-Smith matrix (cf. \cite{Tex16}), we see that \eref{eq:RelationForCorrelationsBetweenProperTimes} corresponds to 
$\sum_{a\neq b}\mean{\tau_a\tau_b}$ and hence provides some information about the correlations between proper times.

These expressions can be used to obtain the statistical properties of the proper time delays.
In the limit of long disordered region, $L\gg\xi$, the fact that $\mean{\tr{\WSm}^2}\simeq\mean{\tr{\WSm^2}}$ shows that proper times are weaky correlated.
Using that all channels are equivalent, we get the mean value
\begin{equation}
  \mean{\tau_a} = \frac{L}{k} = 2\tau_\xi\,\frac{L}{\xi}
  \:,
\end{equation}
the second moment $\mean{\tau_a^2}=(1/\Nc)\mean{\tr{\WSm^2}}$
\begin{equation}
  \mean{\tau_a^2} \simeq \tau_\xi^2\,\frac{\Nc\,\beta}{2(\beta+2)}\,\EXP{4L/\xi}
  \:,
\end{equation}
and from \eref{eq:RelationForCorrelationsBetweenProperTimes}, the correlation 
\begin{equation}
  \mean{\tau_a\tau_b} 
   =
  \tau_\xi^2\,\frac{4}{\beta}\left[\frac{L}{\xi}+\frac{\EXP{-2\beta L/\xi}-1}{2\beta}  \right]
  \simeq \tau_\xi^2\,\frac{4L}{\beta\xi} 
  \hspace{0.5cm}\mbox{for }a\neq b
  \:,
\end{equation}
demonstrating that the proper times are characterized by weak anti-correlation
$\mathrm{Cov}(\tau_a,\tau_b)\simeq-(2\tau_\xi\,L/\xi)^2$.


\section{Resolvent}
\label{sec:MarginalDensity}

\subsection{Resolvent and density of eigenvalues}

We re-examine the problem considered by Ossipov \cite{Oss18}, within our disordered model. 
From the MSDE~\eref{eq:MSDEforQt}, we can deduce an equation for the evolution of the density of eigenvalues of the matrix $\WSmt$, and thus of $\WSm$. 
We introduce the resolvent matrix
\begin{equation}
  \label{eq:defG}
  G(z;L) = \left( z \un - \frac{\Nc}{2\tau_\xi}\, \WSmt(L) \right)^{-1}
  \:.
\end{equation}
The matrix $\WSmt$ is rescaled by a factor $\Nc$, as we expect from the distribution~\eref{eq:LagWishDist} that its eigenvalues scale as $\O(\Nc^{-1})$ for large number of channels.  
From the MSDE~(\ref{eq:MSDEforQt}), we can write the equation satisfied by
$G(z;L)$:
\begin{equation}
  \label{eq:MSDEforG}
  \hspace{-2cm}
  \partial_L G = 
  \frac{N}{\xi} G^2
  - \frac{2\mu}{\xi} G\, (z G - \un)
  + \frac{1}{\sqrt{\xi}} \left[
    (zG-\un)\,\eta(L)\, G + G \,\eta(L)\, (z G - \un)
  \right]
  \:,
\end{equation}
where we have omitted the arguments of $G$ for simplicity. 
Writing $\D B(x) = \eta(x) \D x$ along with the
relations~(\ref{eq:NoiseSquare},\ref{eq:NoiseSquareSandwich}), we can convert
this equation to the It\^o convention. We get
\begin{eqnarray}
  \label{eq:MSDEforGIto}
  \nonumber
  \partial_L G
  =& \frac{N}{\xi} G^2
     - \frac{2\mu}{\xi} G, (z G - \un)
  \\
  \nonumber
   &+ \frac{1}{\xi} (2zG-\un)
     \left[
     -\mu G
     + (2zG-\un) \left(
     \frac{\beta}{2}\tr{G}
     + \left( 1 - \frac{\beta}{2} \right) G
     \right)
     \right]
  \\
   &+ \frac{1}{\sqrt{\xi}} \left[
     (zG-\un)\,\eta(L)\, G + G \,\eta(L)\, (z G - \un)
     \right]
     \hspace{0.5cm}\text{(It\^o)}
     \:.
\end{eqnarray}
We can now take the expectation value, which yields
\begin{eqnarray}
  \nonumber
  \partial_L \mean{G}
  =& \frac{N}{\xi} \mean{G^2}
     - \frac{4\mu z}{\xi} \mean{G^2}
     + \frac{3\mu}{\xi}  \mean{G}
  \\
  \nonumber
   & +\frac{\beta}{2\xi} 
     \left(
     \mean{\tr{G}} -4 z \mean{G \tr{G}} + 4 z^2 \mean{G^2 \tr{G}}
     \right)
  \\
   & +\frac{1}{\xi}
     \left( 1 - \frac{\beta}{2} \right)
     \left(
     \mean{G} -4 z \mean{G^2} + 4 z^2 \mean{G^3}
     \right)
     \:.
\end{eqnarray}
Using that $\partial_z G(z) = -G(z)^2$ and $\partial_z^2 G(z) = 2
G(z)^3$, we can write
\begin{eqnarray}
  \nonumber
  \partial_L \mean{G}
  =& -\frac{N}{\xi} \partial_z \mean{G}
     + \frac{4\mu z}{\xi} \partial_z \mean{G}
     + \frac{3\mu}{\xi}  \mean{G}
  \\
  \nonumber
   & +\frac{\beta}{2\xi} 
     \left(
     \mean{\tr{G}} -4 z \mean{G \tr{G}} - 4 z^2 \mean{(\partial_zG) \tr{G}}
     \right)
  \\
   & +\frac{1}{\xi}
     \left( 1 - \frac{\beta}{2} \right)
     \left(
     \mean{G} +4 z \mean{\partial_z G} + 2 z^2 \mean{\partial_z^2 G}
     \right)
     \:.
\end{eqnarray}
In the limit $N \to \infty$, it is expected that the resolvent
\begin{equation}
  \label{eq:Resol}
  g(z;L) = \frac{1}{\Nc} \tr{G(z;L)}
\end{equation}
becomes deterministic, so we can take it out of the expectation
values. Hence
\begin{eqnarray}
  \nonumber
  \partial_L \mean{G}
  =& -\frac{\Nc}{\xi}  \partial_z \mean{G}
     + \frac{4\mu z}{\xi} \partial_z \mean{G}
     + \frac{3\mu}{\xi}  \mean{G}
  \\
  \nonumber
   & +\frac{\beta}{2\xi} 
     \left(
     \Nc g -4 \Nc z g \mean{G} - 4 \Nc z^2 g \mean{\partial_zG}
     \right)
  \\
   & +\frac{1}{\xi}
     \left( 1 - \frac{\beta}{2} \right)
     \left(
     \mean{G} +4 z \mean{\partial_z G} + 2 z^2 \mean{\partial_z^2 G}
     \right)
     \:.
\end{eqnarray}
Taking the trace of this equation, we can deduce a partial
differential equation on $g(z;L)$:
\begin{equation}
  \label{eq:EvolRes}
  \hspace{-2.5cm}
  \frac{\partial g(z;L)}{\partial L}
  = \frac{1}{\xi} \frac{\partial}{\partial z}\left[
    \Nc \left( 2\left(\beta\,z -\frac12\right)g(z;L)  -\beta z^2 g(z;L)^2 \right)     
    +
    \left( 2 - \beta \right)
    \frac{\partial}{\partial z} (z^2 g(z;L))
  \right]
  \:.
\end{equation}
The last term is subleading in the limit $\Nc \to \infty$ considered here and should be dropped. 
We have kept it here 
to make clear that, for $\beta=1$, Eq.~\eref{eq:EvolRes} exactly coincides with the result of Ref.~\cite{Oss18} (upon rescaling $z\to z/4$ and hence $g\to4g$, due to a different definition). 
The density of eigenvalues $\rho(\lambda;L)$ can then be obtained from $g(z;L)$ as
\begin{equation}
  \rho(\lambda;L) = - \frac{1}{\pi} \im \left[ g(\lambda+\I 0^+;L)\right] 
  \:.
\end{equation}
In Ref.~\cite{Oss18}, a nice solution of Eq.~\eref{eq:EvolRes} was obtained by using the analogy with the Burgers equation.

\subsection{Discussion}

In his article~\cite{Oss18}, Ossipov has claimed that his \og \textit{approach provides (...) foundation for the arguments of the scaling theory of Anderson localization} \fg{}, and that \og \textit{scattering isotropy (...) is not used in our approach} \fg{}, which \og \textit{allows to study the problem in higher dimension} \fg{}.
We disagree with these statements.

For $\beta=1$, one can write $\WSm=\mathcal{O}\,\mathrm{diag}(\tau_1,\cdots,\tau_\Nc)\,\mathcal{O}^\T$, where the orthogonal matrix $\mathcal{O}$ gathers the eigenvectors of the Wigner-Smith matrix.
Ossipov has argued that the derivation of his Eq.~12, i.e. \eref{eq:EvolRes} for $\beta=1$, relies on assumptions that $\Sm$ and $\mathcal{O}$ would be controlled by fast variables, while the eigenvalues $\{\tau_1,\cdots,\tau_\Nc\}$ would be the only slow variables.
First, let us stress that this statement is supported in  Ref.~\cite{Oss18} by its Eq.5, which shows that $\Sm$ and $\mathcal{O}$ are uniformly distributed in the unitary and orthogonal groups, respectively, which is unrelated with the statement of slow/fast variables. 
The correlations between eigenvalues and eigenvectors of $\WSm$ were not investigated in Ref.~\cite{Oss18}.
Second, in our paper, working with $\Smt$ instead of the scattering matrix $\Sm$ has eliminated the fast variables.
This has led to matrix stochastic differential equations for $\Smt$ and the symmetrised Wigner-Smith matrix $\WSmt$.
Using the isotropy assumption, these MSDE have been decoupled, leading to the matrix SDE~\eref{eq:MSDEforQt} for $\WSmt$. Thus we have obtained that the matrix $\WSmt$ is controlled by slow variables, i.e. both its eigenvalues and eigenvectors are slow variables. Their decoupling relies on the isotropy assumption.

The fact that we have recovered Ossipov's equation within a model based on \textit{isotropy} assumption, emphasizes that Ossipov's central equation~12  has the same physical content as the equation obtained within the DMPK approach (Ossipov agrees that his \og \textit{Eq.~8} [from which his Eq.~12 is derived] \textit{coincides with the DMPK equation} \fg{}). 
These equations describe disordered wires which are \textit{transversally ergodic}, i.e. are inherently restricted to the quasi-one-dimensional regime and cannot encode the physics of Anderson localisation in dimension~$d\geq2$.



\section{Conclusion}

We have studied the Wigner-Smith time delay matrix $\WSm$ for multichannel disordered wires of length $L$.
Using an isotropy assumption, we have been able to use the decoupling between fast and slow variables both for the scattering matrix and the Wigner-Smith matrix (the two length scales associated with fast and slow variables are the wavelength $\lambdabar$ and the elastic mean free path $\ell_e$, respectively).
For this purpose, the new symmetrisation of the Wigner-Smith matrix was crucial.
We have provided some matrix stochastic differential equations for matrices controlling the slow variables, which have eventually led to a representation of $\WSmt$, the symmetrised $\WSm$-matrix, under the form of an exponential functional of a matrix Brownian motion (BM). 
This representation is a generalization of the result for $\Nc=1$, obtained by one of us with Comtet \cite{ComTex97,TexCom99}. 

In the limit of semi-infinite disordered region, $L\to\infty$, by making use of an extension of the recent matrix Dufresne identity of Rider and Valk\'o \cite{RidVal15}, we have recovered straightforwardly the distribution of the $\WSm$'s eigenvalues found by Brouwer and Beenakker \cite{Bee01,BeeBro01} by different means.

Furthermore, our exponential functional representation allows to study the statistical properties of $\WSm$ for finite $L$~:
we have derived the first moments.
In particular, we have shown that $\mean{\tr{\WSm^2}}$ and $\mean{\tr{\WSm}^2}$ both behave as $\sim\exp[4L/\xi]$. 
The stucture of the calculation (with the result for $\Nc=1$ of Ref.~\cite{TexCom99}) suggests to conjecture the form 
  \begin{equation}
    \mean{\tr{\WSm^{n_1}}\tr{\WSm^{n_2}}\cdots\tr{\WSm^{n_k}}}
    \sim \EXP{2n(n-1)L/\xi}
    \hspace{0.5cm}\mbox{where }
    n=\sum_{i=1}^k  n_i
    \:.
  \end{equation}
in the general case

In Section~\ref{sec:WSlocalisation}, we have followed an alternative approach for the derivation of the exponential functional representation of the BM. Although this derivation was less rigorous, it emphasizes the universal character of the results. 
Interestingly, it suggests that $\WSm$ has also a representation in terms of exponential functional of the BM. This point still deserves some clarifications.

Finally, our analysis opens the natural question of finding extensions of our results.
Elimination of fast variables relies on a high energy/disorder and some isotropy assumption (invariance between exchange of channels). 
Would it be possible to relax the second hypothesis~?
Is there an exponential functional of the BM representation for non identical channels~? For example by considering the case of channel dependent wave vectors, $k\,\un\to\mathrm{diag}(k_1,\ldots,k_\Nc)$, and/or anisotropic correlations $C_{ab,cd}$, which are relevant to describe more reallistic mutlichannel disordered wires.


\section*{Acknowledgements}

We thank Alain Comtet for stimulating discussions.
This project has benefitted from funding of the Netherlands Organization for Scientific Research (NWO/OCW) and from the European Research Council (ERC) under the European Union’s Horizon 2020 research and innovation programme.


\begin{appendix}

\section{Wigner-Smith matrix in terms of the stationary scattering states~: derivation of Eq.~\eref{eq:MatrixKreinFriedel}}
\label{app:FSrelation}

A relation similar to Eq.~\eref{eq:MatrixKreinFriedel} was derived by Friedel \cite{Fri58} 
and Smith \cite{Smi60} for centro-symmetric potential.
A proof for the more general case of metric graphs was given in Refs.~\cite{TexBut03,TexDeg03} (cf. Eqs.~(43,53) of the first reference, or Eq.~14 of the second).
Here, we briefly adapt the derivation of \cite{Tex02} for metric graphs to the case of multichannel disordered wires.
We consider $\Phi=\sqrt{hv}\,\Psi$, which presents the ``asymptotic'' behaviour  
\begin{equation}
  \label{eq:AsymptoticPhiAppA}
  \Phi(x)  =
  \mathbf{1}_\Nc\,\EXP{-\I k (x-L)} + \Sm(\varepsilon )\, \EXP{\I k (x-L)} 
  \hspace{0.5cm}\mbox{for } x\geq L
  \:.
\end{equation}

\paragraph{Step 1~:}

Let us define the $\Nc\times\Nc$ matrix 
\begin{equation}
  \Omega(x) \eqdef 
  \derivp{\Phi^\dagger}{x}\derivp{\Phi}{\varepsilon}
  - \Phi^\dagger\derivp{^2\Phi}{x\partial\varepsilon}
  \:,
\end{equation}
which satisfies
\begin{equation}
  \label{eq:AppA1}
  \derivp{\Omega(x)}{x}=\Phi^\dagger(x)\Phi(x)
  \:.
\end{equation}

\paragraph{Step 2~:}

We compute $\Omega(x)$ at the boundaries. 
Dirichlet boundary condition gives $\Omega(0)=0$. Using the expression of the scattering state \eref{eq:AsymptoticPhiAppA}, we find
\begin{equation}
  \label{eq:AppA2}
  \Omega(L) = -2\I k \,\Sm^\dagger\derivp{\Sm}{\varepsilon} - \frac{\I}{2k}(\Sm-\Sm^\dagger)
  \:.
\end{equation}

\paragraph{Step 3~:}

We combine
\begin{equation}
  \int_0^L\D x\, \derivp{\Omega(x)}{x} = \Omega(L) - \Omega(0)
\end{equation}
with \eref{eq:AppA1} and \eref{eq:AppA2}.
Multiplication by $1/(4\pi k)$ we end with the matricial identity \eref{eq:MatrixKreinFriedel}.

\subsubsection*{Eq.~\eref{eq:MatrixKreinFriedel} and the DoS.---}

Note that the trace $\rho(x;\varepsilon)=\tr{\Psi_\varepsilon^\dagger(x) \Psi_\varepsilon(x)}$ has the interpretation of the local density of states, thus we recover the Krein-Friedel relation \cite{Tex02,TexBut03,TexDeg03}
\begin{equation}
  \label{eq:TrueKreinFriedel}
  \int_0^L\D x\, \rho(x;\varepsilon)
  = \frac{1}{2\pi}
  \tr{ \WSm }
  + \frac{\tr{\Sm-\Sm^\dagger}}{8\I\pi\varepsilon}
  \:.
\end{equation}


\section{Effective MSDE from the Fokker-Planck equation for $N=2$ and  $\beta=1$}
\label{sec:specCase}

In this Appendix, we show a rigorous procedure to average over the
fast variables based on the Fokker-Planck equation.  Let us consider
for simplicity $N=2$ and $\beta=1$. We take a noise of the form
\begin{equation}
  \label{eq:VN2}
  V =
  \left(\begin{array}{cc}
    \sqrt{\sigma_1} \, v_1 &  \sqrt{\sigma_2/2} \,v_2
    \\
     \sqrt{\sigma_2/2}\,v_2 &  \sqrt{\sigma_3} \, v_3
  \end{array}\right)
  \:,
\end{equation}
where $v_1$, $v_2$ and $v_3$ are independent Gaussian white noises of
unit variance. Setting $\sigma_1 = \sigma_2 = \sigma_3 = \sigma$
corresponds to the isotropic case~(\ref{eq:IsotropyAssumption}).

We parametrise the scattering matrix in terms of its eigenvalues and
eigenvectors
\begin{equation}
  \Sm = \O^\T \: \EXP{2 \I \Phi} \: \O
  \:,
\end{equation}
where
\begin{equation}
  \Phi = \Diag(\phi_1, \phi_2)
\end{equation}
and $\O$ is a rotation matrix:
\begin{equation}
  \O = \left(\begin{array}{cc}
    \cos \theta & -\sin \theta
    \\
    \sin \theta & \cos \theta
  \end{array}\right)
  \:.
\end{equation}
Using this parametrisation in Eq.~(\ref{eq:SDEforS}), we obtain the
equations for $\phi_1$, $\phi_2$ and $\theta$:
\begin{equation}
  \deriv{}{x}
  \left(\begin{array}{cc}
    \phi_1
    \\
    \phi_2
    \\
    \theta
  \end{array}\right)
  = 
  \left(\begin{array}{cc}
    k
    \\
    k
    \\
    0
  \end{array}\right)
  + \frac{1}{k} \, b(\phi_1, \phi_2,\theta) 
  \left(\begin{array}{cc}
    v_1
    \\
    v_2
    \\
    v_3
  \end{array}\right)
  \quad
  \text{(Stratonovich),}
\end{equation}
where $b$ is the following $3\times 3$ matrix
\begin{equation}
  \hspace{-0.5cm}
  \left(\begin{array}{ccc}
          -\sqrt{\sigma_1} \cos^2\theta \cos^2\phi_1
          & \frac{\sqrt{\sigma_2} }{\sqrt{2}} \sin (2\theta ) \cos^2\phi _1
          & - \sqrt{\sigma_3}\sin ^2\theta \cos^2\phi _1
          \\
          - \sqrt{\sigma_1} \sin^2\theta  \cos^2\phi _2
          & -\frac{\sqrt{\sigma_2}}{\sqrt{2}} \sin (2\theta ) \cos^2\phi_2
          & - \sqrt{\sigma_3} \cos^2\theta \cos^2\phi _2
          \\
          \frac{\sqrt{\sigma_1} \cos \phi _1 \cos \phi _2 \sin (2 \theta )}{2 \sin(\phi_1-\phi_2)}
          & \frac{\sqrt{\sigma_2} \cos (2 \theta ) \cos \phi _1 \cos \phi _2}{
            \sqrt{2}\sin (\phi _1-\phi _2) }
          &
            -\sqrt{\sigma_3} \frac{\cos \phi _1 \cos \phi _2 \sin (2 \theta )}{
            2\sin(\phi _1-\phi _2)}
        \end{array}\right)
      \,.
    \end{equation}
Since the two phases evolve rapidly (on the scale $1/k$), we denote
\begin{equation}
  \phi_1 = k x + \tilde{\phi}_1
  \:,
  \quad
  \phi_2 = k x + \tilde{\phi}_2
  \:.
\end{equation}
Our aim is to obtain equations describing the evolution of
$\tilde{\phi}_1$, $\tilde{\phi}_2$, and $\theta$. We will average over
the fast variables $kx$, at the level of the Fokker-Planck equation
\begin{equation}
  \derivp{P}{x}(\tilde{\phi}_1, \tilde{\phi}_2, \theta)
  = \frac{1}{2} \sum_{i,j,l=1}^3 \derivp{}{X_i} b_{il} \derivp{}{X_j} b_{jl} P
  \:,
\end{equation}
where $X=(\tilde{\phi}_1, \tilde{\phi}_2, \theta)$.  The idea is to
average over $kx$ on one period (the slow variables $\tilde{\phi}_1$,
$\tilde{\phi}_2$, and $\theta$ can be considered constant on
this scale). For this, we need all the derivatives to be in the front
(which corresponds to converting the SDEs to It\^o). Then, after
averaging, we obtain an equation of the form
\begin{equation}
  \label{eq:FPEavg}
  \derivp{P}{x}(\tilde{\phi}_1, \tilde{\phi}_2, \theta)
  = \sum_{i,j}^3 \derivp{}{X_i}
  \left( -\tilde{a}_i + \frac{1}{2} \derivp{}{X_j} \: \tilde{c}_{ij} \right) P
  \:,
\end{equation}
which we can interpret as a new Fokker-Planck equation, with a drift
$\tilde{a}$ (in It\^o) and a new matrix $\tilde{c}$ which we need to
decompose into the form $\tilde{c} = \tilde{b} \tilde{b}^\T$ in order
to write the corresponding SDE. To perform this factorisation, let us
consider the contribution of each noise independently. We first set
$\sigma_2=\sigma_3 = 0$ to keep only $v_1$. The resulting matrix
$\tilde{c}$ 
is of rank 3, while the original matrix $b b^\T$ was of rank 1. This
means that the noise $v_1$ gave rise to 3 independent Gaussian white
noises, all controlled by the same variance $\sigma_1$. We can
factorise this matrix for $\sigma_2=\sigma_3=0$ as $\tilde{c} =
\tilde{b}_1 \tilde{b}_1^\T$, where
\begin{equation}
  \hspace{-1cm}
  \tilde{b}_1
  = \frac{\sqrt{\sigma_1}}{2k}
  \left(
    \begin{array}{ccc}
      -\cos^2 \theta
      &-\frac{\cos^2 \theta \cos (2 \phi_1)}{\sqrt{2}}
      &\sqrt{2} \cos^2 \theta \cos \phi_1 \sin \phi_1
      \\
      -\sin^2 \theta
      & -\frac{\sin^2 \theta \cos (2 \phi_2)}{\sqrt{2}}
      & \sqrt{2} \sin^2 \theta \cos \phi_2 \sin \phi_2
      \\
      \frac{\cot(\phi_1-\phi_2) \sin(2 \theta)}{2}
      & \frac{\cos(\phi_1+\phi_2) \sin(2 \theta)}{2 \sqrt{2}
        \sin(\phi_1-\phi_2)}
      & -\frac{\sin(\phi_1+\phi_2) \sin(2 \theta)}{2 \sqrt{2} \sin(\phi_1-\phi_2)}
    \end{array}
  \right)
  \:.
\end{equation}
The same situation occurs for the noises $v_2$ and $v_3$, which each
give rise to three new independent noises. Similarly, we obtain a
matrix $\tilde{b}_2$ for $\sigma_1=\sigma_3=0$ and $\tilde{b}_3$ for
$\sigma_1=\sigma_2=0$. Finally, we can write the full matrix
$\tilde{c} = \tilde{b} \tilde{b}^\T$, where $\tilde{b}$ is the
following $3 \times 9$ matrix, with block structure
\begin{equation}
  \tilde{b} =
  \left(
    \begin{array}{ccc}
      \tilde{b}_1& \tilde{b}_2 & \tilde{b}_3
    \end{array}
  \right)
  \:.
\end{equation}
We can then rewrite the Fokker-Planck equation~(\ref{eq:FPEavg}) in
the form
\begin{equation}
  \derivp{P}{x}(\tilde{\phi}_1, \tilde{\phi}_2, \theta)
  = \frac{1}{2} \sum_{i,j=1}^3 \sum_{p=1}^9 \derivp{}{X_i}
  \left( \tilde{b}_{ip} \derivp{}{X_j} \: \tilde{b}_{jp} \right) P
  \:,
\end{equation}
where the drift terms have cancelled out with the terms coming from
the derivative of the matrix $\tilde{b}$. We can write the
corresponding SDE by introducing three new independent matrices of
white noises, which we denote $\tilde{V}_i$, $i=1,2,3$. Rewriting the
result in terms of
\begin{equation}
  \Smt = \O^\T \: \Diag(\EXP{2 \I \tilde{\phi}_1}, \EXP{2 \I \tilde{\phi}_2}) \: \O
  = \EXP{-2\I k x} \: \Sm
\end{equation}
gives
\begin{equation}
  \label{eq:EvolStilde}
  \hspace{-2cm}
    \partial_x \Smt = \frac{1}{2\I k}
    \left\lbrace
      V_1 + \Smt V_1 \Smt
      - \I
      \left(V_2 - \Smt V_2 \Smt \right)
      + V_3 \Smt + \Smt V_3
    \right\rbrace
    \:,
    \quad \text{(Stratonovich)}
\end{equation}
where
\begin{equation}
  V_1 \eqlaw V_2 \eqlaw \frac{1}{\sqrt{2}} V
  \:,
  \quad
  V_3 \eqlaw V
  \:,
  \quad
  V_i \text{ and } V_j \text{ independent.}
\end{equation}

\end{appendix}

\section*{References}
%

\end{document}